\documentclass{aa}
\usepackage{amssymb,astron,times,graphics}
\begin{document}
\thesaurus{06(08.16.4; 08.03.1; 08.03.4; 08.12.1; 13.19.5)}
\title{Models of circumstellar molecular radio line emission\thanks{Presented
in this paper is observational data collected using the
Swedish-ESO submillimetre telescope, La Silla, Chile,
the 20\,m telescope at Onsala Space Observatory, Chalmers Tekniska H\"ogskola,
Sweden, and the NRAO 12\,m telescope located at Kitt Peak, USA.}}
\subtitle{Mass loss rates for a sample of bright carbon stars}

\author{F.~L.~Sch\"oier\inst{1} \and H.~Olofsson\inst{1}}

\institute{Stockholm Observatory, SE-133 36 Saltsj\"obaden, Sweden}

\offprints{F.~L.~Sch\"oier (fredrik@strw.leidenuniv.nl)}
\date{A\&A in press}

\maketitle
\begin{abstract}
Using a detailed radiative transfer analysis, combined with an
energy balance equation for the gas, we have performed extensive
modelling of circumstellar CO radio line emission from a large sample
of optically bright carbon stars, originally observed by Olofsson et
al.  (ApJS, 87, 267).  Some new observational results are presented
here. We determine some of the basic parameters that characterize
circumstellar envelopes (CSEs), e.g., the stellar mass loss rate, the
gas expansion velocity, and the kinetic temperature structure of the
gas.  Assuming a spherically symmetric CSE with a smooth gas density
distribution, created by a continuous mass loss, which expands with a
constant velocity we are able to model reasonably well 61 of our 69
sample stars.  The derived mass loss rates depend crucially on the
assumptions in the circumstellar model, of which some can be
constrained if enough observational data exist.  Therefore, a reliable
mass loss rate determination for an individual star requires, in
addition to a detailed radiative transfer analysis, good observational
constraints in the form of multi-line observations and radial
brightness distributions.  In our analysis we use the results of a
model for the photodissociation of circumstellar CO by Mamon et al.\
(1988).  This leads to model fits to observed radial brightness
profiles that are, in general, very good, but there are also a few
cases with clear deviations, which suggest departures from a simple
$r^{-2}$ density law.

The derived mass loss rates span almost four orders of magnitude, from
$\sim$5$\times$10$^{-9}$\,M$_{\sun}$\,yr$^{-1}$ up to
$\sim$2$\times$10$^{-5}$\,M$_{\sun}$\,yr$^{-1}$, with the median mass
loss rate being 2.8$\times$10$^{-7}$\,M$_{\sun}$\,yr$^{-1}$.
We estimate that the mass loss rates are typically accurate to
$\sim$50\% within the adopted circumstellar model. The physical conditions
prevailing in the CSEs vary considerably over such a large range of
mass loss rates.  Among other things, it appears that the dust-to-gas
mass ratio and/or the dust properties change with the mass loss rate.
We find that the mass loss rate and the gas expansion velocity are
well correlated, and that both of them clearly depend on the pulsational
period and (with larger scatter) the stellar luminosity.  Moreover, the mass
loss rate correlates weakly with the stellar effective temperature, in
the sense that the cooler stars tend to have higher mass loss rates,
but there seems to be no correlation with the stellar C/O-ratio.
We conclude that the mass loss rate increases with increased
regular pulsation and/or luminosity, and that the expansion velocity
increases as an effect of increasing mass loss rate (for low mass
loss rates) and luminosity.

Five, of the remaining eight, sample stars have detached CSEs in the form of
geometrically thin CO shells.  The present mass loss rates and shell
masses of these sources are estimated.  Finally, in three cases we encounter
problems using our model.  For two of these sources there are
indications of significant departures from overall spherical symmetry
of the CSEs.

Carbon stars on the AGB are probably important in returning processed gas to
the ISM. We estimate that carbon stars of the type considered here
annually return $\sim$0.05\,M$_{\sun}$ of gas to the Galaxy, but more
extreme carbon stars may contribute an order of magnitude more.
However, as for the total carbon budget of the Galaxy, carbon stars
appear to be of only minor importance.

\end{abstract}

\section{Introduction}

The carbon star phenomenon, i.e., the presence of stars for which the
abundance of carbon exceeds that of oxygen in the atmospheres, occurs during
the final evolutionary stage of low to intermediate mass stars
($\sim$1$-$10\,M$_{\sun}$; Wallerstein \& Knapp 1998\nocite{WK}), and it is
believed to be due to a dredge-up process driven by quasi-periodic
He-shell flashes (Straniero et al.\ 1997\nocite{Straniero97}).
These stars, located on
the asymptotic giant branch (AGB), lose copious amounts of matter
($\sim$10$^{-8}$$-$10$^{-4}$\,M$_{\sun}$\,yr$^{-1}$) in an intense
stellar wind.  In fact, it is the mass loss that determines the AGB
lifetime and not the nuclear burning processes.  Hence, a
determination of the mass loss characteristics, i.e., the dependence
on time, mass, chemistry, metallicity, etc., is of utmost importance
for our understanding of late stellar evolution.

The mass loss creates a circumstellar envelope (CSE) of gas and dust around
the star.  The low temperature of the central star allows the
formation of a wide variety of molecular species in its atmosphere, and
the expanding gas is therefore mainly in molecular form.  The chemistry in the
CSE itself can be very rich, and it depends on the C/O-ratio, the
thickness of the envelope, and the strength of the ambient ultraviolet
radiation field.  Presently, $\sim$60 different molecular species have been
identified in CSEs around AGB-stars (Olofsson 1997\nocite{Olofsson97}).

The study of AGB-CSEs is important for the understanding of the late
stages of stellar evolution as mentioned above.  In addition, an
understanding of the atmospheric and circumstellar chemistry is
required to determine the elemental abundances in these winds, which
contribute to the chemical evolution of the interstellar medium (Busso
et~al.\ 1999\nocite{Busso99}).  Finally, since AGB stars are the
progenitors of planetary nebulae (PNe), the CSEs are very likely
important ingredients in the formation of PNe (Kwok 1993\nocite{Kwok93}).

Observations of
molecular millimetre-wave line emission have proven to be one of the
best tools for studying the structure, kinematics, and chemistry of
the CSEs, and CO and OH observations have proven particularly useful
for the determination of accurate mass loss rates
[see Olofsson \cite*{Olofsson96a} and references therein].

While the overall picture of mass loss on the AGB is fairly
well understood, some of the underlying physical principles are not.
The prevailing theory, which needs to be thoroughly tested
observationally, is that the mass loss occurs in a two stage process.
First the pulsations of the star deposit energy in the atmosphere,
leading to a considerably increased scale height, and sufficient
matter at low enough temperatures for efficient dust formation.  Radiation
pressure on the dust grains accelerate the dust wind to its terminal
expansion velocity in the second stage.  The gas is momentum
coupled, through collisions, to the dust, and it will be effectively
dragged along.  The mechanisms that cause the mass loss to vary on a
wide range of time scales still need to be pinpointed.  Eventually, the
possibility of mass ejected in clumps has to be addressed,
although in this paper we will consider only a smooth spherically
symmetric wind.

This study of an essentially complete sample of visually bright, relatively
unobscured carbon stars will hopefully increase our knowledge of the
modelling of circumstellar molecular radio line emission, the mass
loss characteristics, the circumstellar chemistry, the elemental
abundances in the winds, and possibly the evolutionary status of these
objects.  It will also provide us with the tools required to determine
the properties of higher mass loss rate objects, for which the central
star may be completely obscured.

In this paper the basic physical characteristics of the CSEs, e.g., the
stellar mass loss rate, the gas kinetic temperature, and the gas expansion
velocity, are determined using a radiative transfer analysis of the
observed $^{12}$CO (hereafter CO) millimetre-wave line emission.  The
CO molecule is particularly well suited for this purpose, since it is
difficult to photodissociate and easy to excite through collisions,
and thus is a very good tracer of the molecular gas density and
temperature.
Furthermore, the molecular excitation is of a quasi-thermal nature which
simplifies detailed modelling.

\begin{table}
  \caption{Data on telescopes and receivers used.}
  \label{teldat}
  \resizebox{\hsize}{!}{
  \begin{tabular}{llccccc} \hline
  \noalign{\smallskip}
  \multicolumn{1}{c}{Tel.}    & 
  \multicolumn{1}{c}{Trans.}    & 
  \multicolumn{1}{c}{$\nu$}    & 
  \multicolumn{1}{c}{$T_{\mathrm{rec}}$(SSB)}    & 
  \multicolumn{1}{c}{$\theta_{\mathrm{mb}}$}    &  
  \multicolumn{1}{c}{$\eta_{\mathrm{mb}}$} &
  \multicolumn{1}{c}{$\eta^*_{\mathrm{m}}$} \\
  & 
  &
  \multicolumn{1}{c}{[GHz]} &
  \multicolumn{1}{c}{[K]} &
  \multicolumn{1}{c}{[$\arcsec$]} & \\
  \noalign{\smallskip}
  \hline
  \noalign{\smallskip}
  OSO  & CO(1$-$0) & 115.271 &           100 & 33 & 0.5\phantom{0} &      \\
  SEST & CO(1$-$0) & 115.271 &           110 & 45 & 0.7\phantom{0} &      \\
       & CO(2$-$1) & 230.538 &           110 & 23 & 0.5\phantom{0} &      \\
       & CO(3$-$2) & 345.796 &           300 & 16 & 0.25           &      \\  
  NRAO & CO(1$-$0) & 115.271 & \phantom{0}80 & 55 & 0.55           & 0.84 \\ 
       & CO(2$-$1) & 230.538 &           200 & 27 & 0.30           & 0.45 \\      
  \hline
  \noalign{\smallskip}
  \end{tabular}}
\end{table}

\section{Observations}

\subsection{The sample}

We have selected a sample of 68 bright N- and J-type
carbon stars for which
circumstellar CO emission was detected by Olofsson et~al.\ (1993a).
Their original sample consisted of carbon stars brighter than
$K$$=$2\,mag., in total 120 sources.  The detection rate for
all the carbon stars searched for circumstellar CO was as high as
72\%.  The original sample is thought to be close to complete to
distances below 1\,kpc [it is estimated that about a third of the
carbon stars within this distance are missing, presumably the ones
with thick and dusty CSEs, i.e., the high mass loss objects (Olofsson
et~al.  1993a)], and for distances below 500\,pc all sources (a total
of 41) were detected in CO (note that the distances used in this paper
are generally lower than those adopted in Olofsson et~al.\
1993a\nocite{Olofsson93a}).  These facts taken together led to the
conclusion that the great majority of all N-type carbon stars are losing mass
at a rate higher than about 10$^{-8}$\,M$_{\sun}$\,yr$^{-1}$.  They do
not, however, constitute a homogeneous group in terms of circumstellar
properties.  Instead they show a wide range of gas expansion
velocities and mass loss rates, presumably due to differences in mass,
evolutionary stage, etc..  In addition, the reasonably well studied,
high mass loss rate carbon star, LP And (also known as IRC+40540), is
included in our analysis, since it provides a good test case for the
radiative transfer model.

The CO ($J$$=$$1$$\rightarrow$$0$ and $J$$=$$2$$\rightarrow$$1$)
observations presented in Olofsson et~al.  \cite*{Olofsson93a} were
obtained using the 20\,m telescope at Onsala Space Observatory (OSO),
Sweden, the 15\,m Swedish-ESO sub-millimetre telescope (SEST) at La
Silla, Chile, and the IRAM 30\,m telescope at Pico Veleta, Spain,
during the years 1986$-$1992.

\begin{figure*}
        \resizebox{17cm}{!}{\includegraphics{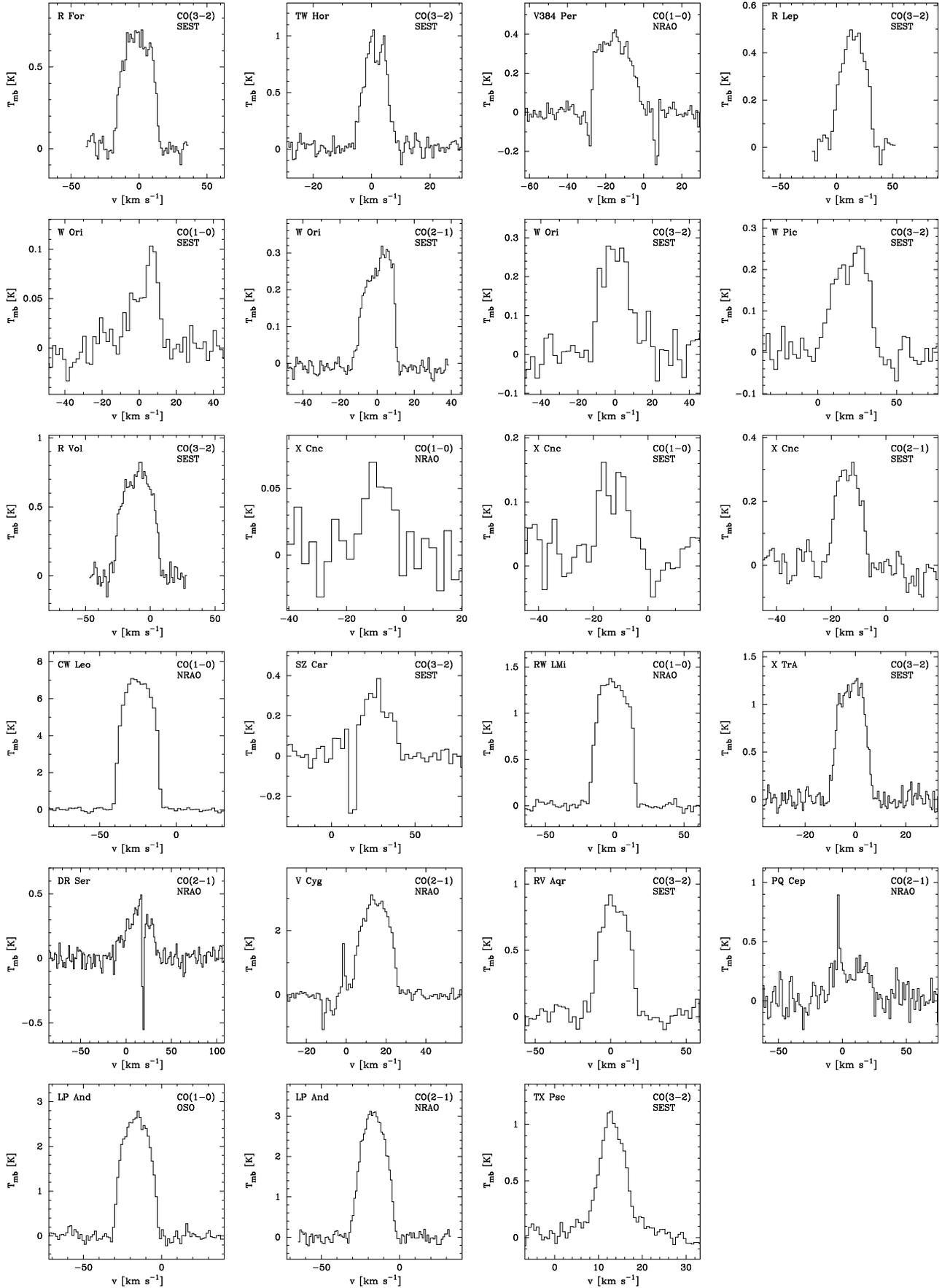}}
        \caption{New observations of circumstellar
        $^{12}$CO line emission.  The line observed and the telescope used are
        shown in the upper right corner of each panel.}
        \label{spectra}
\end{figure*}

\begin{table*}
  \caption{Observational results (see text for details). The intensity scale is
  thought to be accurate to within $\sim$20\%.}
  \label{obs}
  \begin{tabular}{llcccccc} \hline
  \noalign{\smallskip}
  \multicolumn{1}{c}{Source}    & 
  \multicolumn{1}{c}{Tel.}    & 
  \multicolumn{1}{c}{Trans.}    & 
  \multicolumn{1}{c}{$T_{\mathrm{mb}}$}    & 
  \multicolumn{1}{c}{$I_{\mathrm{mb}}$}    &  
  \multicolumn{1}{c}{$v_*$}    & 
  \multicolumn{1}{c}{$v_\mathrm{e}$} &
  \multicolumn{1}{c}{$\beta$} \\
  & 
  &
  &
  \multicolumn{1}{c}{[K]} &
  \multicolumn{1}{c}{[K\,km\,s$^{-1}$]} &
  \multicolumn{1}{c}{[km\,s$^{-1}$]} &
  \multicolumn{1}{c}{[km\,s$^{-1}$]} & \\
  \noalign{\smallskip}
  \hline
  \noalign{\smallskip}
   \object{R For}      & SEST & 3$-$2 & 0.70           &\phantom{0}17.9 &\phantom{0}$-$1.8 &	     16.1 & \phantom{$-$}1.0  	  \\
   \object{TW Hor}     & SEST & 3$-$2 & 0.94           &\phantom{00}8.7 &\phantom{$-$0}1.0 &\phantom{0}7.5 & \phantom{$-$}2.5 \\	
  \object{V384 Per}$^1$& NRAO & 1$-$0 & 0.37           &\phantom{00}7.8 &          $-$16.4 &          11.5 & \phantom{$-$}0.5     \\	   
   \object{R Lep}      & SEST & 3$-$2 & 0.48           &\phantom{0}12.9 &\phantom{$-$}12.4 &	     19.5 & \phantom{$-$}1.7\\
   \object{W Ori}$^1$  & SEST & 1$-$0 & 0.06           &\phantom{00}1.2 &\phantom{0$-$}1.6 &\phantom{0}8.6 & $-$0.7\\
                       & SEST & 2$-$1 & 0.27           &\phantom{00}4.9 &\phantom{0$-$}0.0 &	     10.0 & \phantom{$-$}0.4\\
                       & SEST & 3$-$2 & 0.27           &\phantom{00}4.8 &\phantom{0}$-$1.4 &	     11.8 & \phantom{$-$}1.1\\
   \object{W Pic}      & SEST & 3$-$2 & 0.24           &\phantom{00}6.3 &\phantom{$-$}20.7 &	     13.9 & \phantom{$-$}0.4    	  \\
   \object{R Vol}      & SEST & 3$-$2 & 0.74           &\phantom{0}20.8 &  	   $-$10.6 &	     16.9 & \phantom{$-$}0.8          \\
   \object{X Cnc}      & NRAO & 1$-$0 & 0.06           &\phantom{00}0.7 &  	   $-$10.3 &\phantom{0}7.6 & \phantom{$-$}1.3          \\
                       & SEST & 1$-$0 & 0.12           &\phantom{00}1.7 &  	   $-$12.8 &\phantom{0}8.6 & \phantom{$-$}1.0          \\
                       & SEST & 2$-$1 & 0.30           &\phantom{00}3.2 &  	   $-$14.3 &\phantom{0}7.5 & \phantom{$-$}1.7          \\
   \object{CW Leo}     & NRAO & 1$-$0 & 7.1\phantom{0} &          170.8 &  	   $-$25.9 &         14.3 & \phantom{$-$}0.7          \\
   \object{SZ Car}$^1$ & SEST & 3$-$2 & 0.30           &\phantom{00}5.5 &\phantom{$-$}25.7 &	     13.1 & \phantom{$-$}1.2\\
   \object{RW LMi}     & NRAO & 1$-$0 & 1.3\phantom{0} &\phantom{0}36.4 &\phantom{0}$-$1.6 &         15.8 & \phantom{$-$}0.7          \\
   \object{X TrA}      & SEST & 3$-$2 & 1.2\phantom{0} &\phantom{0}15.3 &\phantom{0}$-$2.0 &\phantom{0}8.7 & \phantom{$-$}1.6\\
   \object{DR Ser}$^1$ & NRAO & 2$-$1 & 0.26           &\phantom{0}11.3 &\phantom{$-$}11.6 &	     24.1 & \phantom{$-$}1.1\\
   \object{V Cyg}      & NRAO & 2$-$1 & 2.9\phantom{0} &\phantom{0}50.1 &\phantom{$-$}14.2 &	     11.0 & \phantom{$-$}1.1\\
   \object{RV Aqr}     & SEST & 3$-$2 & 0.83           &\phantom{0}18.6 &\phantom{$-$0}1.2 &	     14.5 & \phantom{$-$}1.1\\
   \object{PQ Cep}$^1$ & NRAO & 2$-$1 & 0.28           &\phantom{00}8.7 &\phantom{$-$0}6.3 &	     24.1 & \phantom{$-$}0.0 \\
   \object{LP And}     & OSO  & 1$-$0 & 2.6\phantom{0} &\phantom{0}57.0 &          $-$16.9 &	     13.8 & \phantom{$-$}1.1\\
                       & NRAO & 2$-$1 & 3.1\phantom{0} &\phantom{0}60.7 &          $-$17.1 &	     14.0 & \phantom{$-$}1.7\\
   \object{TX Psc}$^2$ & SEST & 3$-$2 & 1.0\phantom{0} &\phantom{00}7.0 &\phantom{$-$0}1.0 &\phantom{0}7.5 & \phantom{$-$}2.0\\
  \hline
  \noalign{\smallskip}
  \noalign{$^1$Contamination by interstellar lines. $^2$$\beta$ was fixed.}
  \end{tabular}
\end{table*}

\subsection{New observations}

We have supplemented the original data with new CO observations of some of the
sample stars.  In particular, in the $J$$=$$3$$\rightarrow$$2$ line using
the SEST (August 1992).  Also some $J$$=$$1$$\rightarrow$$0$ and
$J$$=$$2$$\rightarrow$$1$ data were obtained at SEST (October 1998).
A few sources were observed in the $J$$=$$1$$\rightarrow$$0$ and
$J$$=$$2$$\rightarrow$$1$ lines with the NRAO 12\,m telescope at Kitt
Peak, USA (June 1998), and some $J$$=$$1$$\rightarrow$$0$ data were
obtained using OSO (1998 to 1999).

A summary of telescope and receiver data [the receiver temperature
$T_{\mathrm{rec}}$ in single sideband mode (SSB), the full half power
main beam width $\theta_{\mathrm{mb}}$, the main beam efficiency
$\eta_{\mathrm{mb}}$, and the corrected main beam efficiency
$\eta^*_{\mathrm{m}}$] at the observational frequencies are given in
Table~\ref{teldat}.

At SEST, a dual channel SIS mixer receiver was used to simultaneously
observe at 115\,GHz (the $J$$=$$1$$\rightarrow$$0$ line) and 230\,GHz
(the $J$$=$$2$$\rightarrow$$1$ line).  At 345\,GHz (the
$J$$=$$3$$\rightarrow$$2$ line), a single polarization SiS mixer
receiver was used.  As backends, two acousto-optical spectrometers
(AOS) were used; one high resolution spectrometer (HRS) and one low
resolution spectrometer (LRS).  The wideband (1\,GHz) LRS had 1440
channels separated by 0.7\,MHz, whereas the narrow band (84\,MHz) HRS
used 2000 channels separated by 42\,kHz.

For the NRAO 12\,m observations we used a dual polarization SIS receiver. As
backends two filterbanks, each with 256 channels and a channel separation
of 1\,MHz, were used.
The spectra obtained at each of the polarizations
were added to increase the signal-to-noise ratio.  In addition, for
strong sources, a millimetre autocorrelator (MAC) was used in order to
obtain a higher velocity resolution.  The MAC was used in a
configuration where the usable bandwith was 300\,MHz and the
resolution was 98\,kHz.

At OSO, a single polarization SIS receiver was used for the observations.  As
backends two filterbanks, with bandwidths of 512\,MHz (MUL~A) and
64\,MHz (MUL~B), were used.  The MUL~A used 512 channels separated by
1\,MHz, and the MUL~B filterbank used 256 channels with a separation of
250\,kHz.

The SEST and OSO observations were made in a dual beamswitch mode,
where the source is alternately placed in the signal and the reference
beam, using a beam throw of about 12$\arcmin$ at SEST and 11$\arcmin$
at OSO. This method produces very flat baselines.  The intensity
scales are given in main beam brightness temperature,
$T_{\mathrm{mb}}$.  $T_{\mathrm{mb}}$$=$$T_{\mathrm
A}^{*}/\eta_{\mathrm{mb}}$, where $T_{\mathrm A}^{\star}$ is the
antenna temperature corrected for atmospheric attenuation using the
chopper wheel method, and $\eta_{\mathrm{mb}}$ is the main beam
efficiency, see Table~\ref{teldat}.

At the NRAO 12\,m telescope the observations were carried out using a
position switching mode, with the reference position located $+$10$\arcmin$ in
azimuth. This is the preferred observation mode for spectral line
observations at the 12\,m telescope. The raw spectra, which are stored in the
$T^*_{\mathrm{R}}$ scale, were converted using
$T_{\mathrm{mb}}$$=$$T_{\mathrm R}^{*}/\eta^*_{\mathrm{m}}$, where
$\eta^*_{\mathrm{m}}$ is the corrected main beam efficiency given in
Table~\ref{teldat}. The $T^*_{\mathrm{R}}$ scale is related to
$T_{\mathrm A}^{\star}$ through
$T^*_{\mathrm{R}}$$=$$T_{\mathrm A}^{\star}$/$\eta_{\mathrm{fss}}$, where
$\eta_{\mathrm{fss}}$ is the forward scattering and spillover efficiency.

Regular pointing checks were made on strong SiO masers (OSO, SEST) and
strong continuum sources (NRAO).  The pointing was usually consistent
within $\sim$3$\arcsec$ for SEST and OSO and $\sim$5$\arcsec$ at NRAO.
The uncertainties in the absolute intensity scales at the various
telescopes are estimated to be about $\pm$15$-$20\%.  However, due to the
low efficiency of the SEST at 345\,GHz, and the narrow beam, we
consider these data to be particularly uncertain in terms of
calibration.

The new observational results are summarized in Table~\ref{obs} and the
spectra are shown in Fig.~\ref{spectra}.  The line parameters, i.e.,
the main beam brightness temperature at the line centre
($T_{\mathrm{mb}}$), the line centre velocity ($v_{*}$), and half the
full line width ($v_{\mathrm e}$), are obtained by fitting the
following line profile to the data (Olofsson et~al.\ 1993a)
\begin{equation}
T(v) = T_{\mathrm{mb}} \left[ 1 - \left( \frac{v-v_*}{v_{\mathrm e}} \right) ^2
\right] ^{\beta/2},
\end{equation}
where $\beta$ is a parameter describing the shape of the line
($\beta$=2 represents a parabolic line shape, and $\beta$$<$0 means a
profile with horns at the extreme velocities).  The integrated intensity
($I_{\mathrm{mb}}$) is obtained by integrating the emission between
$v_{*}\pm$$v_{\mathrm e}$.

In addition, we have obtained publicly available data from the James
Clerk Maxwell Telescope (JCMT) at Mauna Kea, Hawaii.  The JCMT data
are taken at face value.  However, in the cases where there are more
than one observation available, the derived line intensities are
generally consistent within $\pm$20\%.  The good agreement with
corresponding SEST intensities further support the reliability of the JCMT
public data.  Furthermore, interferometer observations of the
CO($J$=$1$$\rightarrow$$0$) brightness distribution around some of our
carbon stars have been obtained by Neri et~al.\ \cite*{Neri} using the
IRAM Plateau de Bure interferometer (PdBI).  The data are publicly
available and have been used in this paper.

\section{Radiative transfer}

\subsection{The Monte Carlo method}
In order to model the circumstellar molecular line emission and to derive
the basic characteristics of the CSEs we have developed a non-LTE
radiative transfer code based on the Monte Carlo method
[Bernes \cite*{Bernes79}; see also Sch\"{o}ier \cite*{Schoier00a}, 
for details].
An accurate treatment of the molecular excitation is needed for a correct
description of the radiative transfer in an expanding circumstellar envelope,
e.g., the Sobolev approximation has been shown to introduce significant errors
in these circumstances (e.g., Sch{\"o}nberg 1985\nocite{Schonberg85}).
The Monte Carlo method is well suited for the study of CSEs, since it is very
flexible and close to the physics and yet simple in principle.  For
instance, one may include very complex geometries and velocity fields
without being forced away from the physics of the problem.  Crosas \&
Menten 1997\nocite{Crosas97} have recently used the Monte Carlo method
to model the circumstellar CO radio line emission of the prominent
carbon star IRC+10216 (from hereon called CW~Leo).

The aim of the radiative transfer is to obtain the steady-state level
populations, of the molecule under study, using the statistical
equilibrium equations (SE). In the Monte Carlo method information on the
radiation field is obtained by simulating the line photons using a
number of model photons, each representing a large number of real
photons from all transitions considered.  These model photons, emitted both
locally in the gas as well as injected from the boundaries of the CSE,
are followed through the CSE and the number of absorptions are
calculated and stored.  Photons are spontaneously emitted in the gas
with complete angular and frequency redistribution (CRD), i.e., the
local emission is assumed to be isotropic and the scattering are
assumed to be incoherent.  The weight of a model photon is continuously
modified, to take the absorptions and stimulated emissions into
account, as it travels through the CSE. When all model photons are
absorbed in, or have escaped, the CSE the SE are solved and the whole
process is then repeated until some criterion for convergence is fulfilled.
Once the molecular excitation, i.e., the level populations, is obtained
the radiative transfer equation can be solved exactly.

The main drawback of the Monte Carlo method is its slow convergence
($\sim$$\sqrt{N_\mathrm{iter}}$).  This is, however, usually
outweighed by its great adaptability.  One of our motivations for
choosing the Monte Carlo method was to be able to treat varying mass
loss rates, departures from spherical symmetry, and the possibility of
a highly clumped medium.  These complicating issues will be
investigated in a future version of our Monte Carlo code.

\subsection{The standard model}

The numerical modelling of line emission from CSEs, where intricate
interplays between physical and chemical processes take place, is a
challenge.  This applies, in particular, to the important inner
regions of the CSE model, where very few observational constraints are
available and our knowledge is very limited.  In this analysis, we
will neglect many of these complexities and assume a relatively simple, yet
reasonably realistic, CSE model. By applying the model to a large sample we
will derive fairly reliable mass loss rates, and we will also be able
to pick out objects for which there appears to be clear deviations
from the simple picture.

In what will be referred to as "the standard model" we assume a spherically
symmetric CSE that expands at a constant velocity.
In our data we see no evidence for gas acceleration in the CO emitting region.
The CSE is divided into
a number of discrete shells, each with its own set of physical
parameters describing the state of the molecular gas.  The density
structure (as a function of distance, $r$, from the central star) can
then be derived from the conservation of mass
\begin{equation}
\rho_{\mathrm{H}_2} = n_{\mathrm{H}_2} m_{\mathrm{H}_2} =
\frac{\dot{M}}{4\pi r^2 v_{\mathrm e}},
\end{equation}
where $\dot{M}$ is the hydrogen gas mass loss rate, and $v_{\mathrm
e}$ is the gas expansion velocity taken from the line profile fits of
Olofsson et~al.  (1993a) or Table~\ref{obs}.  We assume that in the
CSEs of interest here the hydrogen is in molecular form in the region
probed by the CO emission (Glassgold \& Huggins, 1983\nocite{Glassgold83}).
The turbulent
velocity is assumed to be equal to 0.5\,km\,s$^{-1}$ throughout the
entire CSE.

The excitation of the CO molecules were calculated taking into account
30 rotational levels in each of the ground and first vibrational
states.  The radiative transition probabilities and energy levels are taken
from Chandra et~al.  \cite*{Chandra96}, and the rotational collisional rate
coefficients (CO-H$_2$) are based on the results in Flower \& Launay
\cite*{Flower85}.  These are further extrapolated for $J$$>$11 and for
temperatures higher than 250\,K.  Collisional transitions between
vibrational levels are not important due to the low densities and their short
radiative lifetimes.

The kinetic gas temperature is calculated after each iteration in a 
self-consistent way, 
using the derived level populations, by solving the energy balance
equation (e.g., Goldreich \& Scoville 1976)\nocite{Goldreich76},
\begin{equation}
\frac{dT}{dr} = (2-2\gamma)\frac{T}{r} + \frac{\gamma-1}
{n_{\mathrm H_2} k v_{\mathrm e}} (H-C),
\label{ebal}
\end{equation}
where $\gamma$ is the adiabatic index [assumed to be equal to 5/3,
but the results are not very sensitive to its exact value, see also
Groenewegen \cite*{Groenewegen94} and Ryde et al.  \cite*{Ryde99}],
$k$ is the Boltzmann constant, $H$ is the total heating rate per unit
volume, and $C$ is the total cooling rate per unit volume.  The first
term on the right hand side is the cooling due to the adiabatic
expansion of the gas.  Additional cooling is provided by molecular
line emission from CO and H$_2$.  The molecular cooling due to CO is
calculated from the derived level populations using the expression of
Crosas \& Menten \cite*{Crosas97}.  For the H$_2$ cooling we use the
approach by Groenewegen \cite*{Groenewegen94}.  HCN could be an
important coolant in the inner parts of the envelope (Cernicharo
et~al.  1996)\nocite{Cernicharo96}, but Groenewegen
\cite*{Groenewegen94} demonstrates that HCN cooling is only of minor
importance.  In the present version of the code HCN cooling is
therefore not included (in a forthcoming paper, Sch\"{o}ier \&
Olofsson in prep., this issue will be addressed).

The mechanism responsible for the observed mass loss is probably (at
least for the stars of interest here) radiation pressure acting on
small dust grains, which in turn are coupled to the gas
(e.g., H\"{o}fner 1999\nocite{Hoefner99}).  The radiation
pressure on the dust grains will give them a drift velocity,
$v_{\mathrm{dr}}$, relative to the gas (Gilman 1972\nocite{Gilman72};
Goldreich \& Scoville 1976\nocite{Goldreich76}; Kwan \& Hill
1977\nocite{Kwan77})
\begin{equation}
v_{\mathrm{dr}} = \left( \frac{L\,v_{\mathrm e}\,Q}{\dot{M} c}
                        \right) ^{1/2},
\label{v_dr}
\end{equation}
where $L$ is the luminosity, $Q$ is the averaged momentum transfer efficiency,
and $c$ is the speed of light [see Sahai (1990)\nocite{Sahai90} for a
more elaborate treatment].  As a result of the dust-gas drift, kinetic
energy of the order of $\frac{1}{2} m_{\mathrm{H_2}} v^2_{\mathrm{dr}}$
will be transfered to the gas each time a particle collides with a
dust grain.  This is assumed to provide the dominating heating of the
gas (Goldreich \& Scoville 1976\nocite{Goldreich76}; Kwan \& Hill
1977\nocite{Kwan77}),
\begin{equation}
H_{\mathrm{dg}} = (n_{\mathrm{d}} \sigma_{\mathrm{d}} v_{\mathrm{dr}})
\frac{1}{2} \rho_{\mathrm{H_2}} v_{\mathrm{dr}}^2,
\label{H_dg1}
\end{equation}
where $n_{\mathrm{d}}$ is the number density of dust grains, and
$\sigma_{\mathrm{d}}$ the geometrical cross section of a dust grain.
Introducing the dust-to-gas mass ratio $\Psi$ this equation can be written
\begin{equation}
H_{\mathrm{dg}} = \frac{3}{8} m_{\mathrm{H_2}}^2 n_{\mathrm{H_2}}^2
                        \frac{\Psi}{a_{\mathrm d} \rho_{\mathrm d}}
		  \frac{v_{\mathrm{dr}}^3}{1+\frac{v_{\mathrm{dr}}}
		  {v_\mathrm{e}}},
\label{H_dg}
\end{equation}
where $a_{\mathrm d}$ and $\rho_{\mathrm d}$ are the average size and
density of a dust grain, respectively.

When solving the energy balance equation free parameters describing
the dust, i.e., via the dust-gas collisional heating, are introduced.
These are highly uncertain, but affect the derived line intensities.
Here we assume that the $Q$-parameter, i.e., the efficiency of
momentum transfer, to be equal to 0.03
[for details see Habing et~al.\ (1994)\nocite{Habing94}],
and define a new parameter that contains the other dust parameters,
\begin{equation}
h = \left(\frac{\Psi}{0.01}\right)
\left(\frac{2.0\,\mathrm{g\,cm}^{-3}}{\rho_\mathrm{d}}\right)
\left(\frac{0.05\,\mu\mbox{m}}{a_\mathrm{d}}\right),
\label{h}
\end{equation}
where the normalized values are the ones we used to fit the CO line
emission of \object{CW~Leo} using our model, i.e., $h$$=$1 for this object.

Additional heating is provided by the photoelectric effect
(Huggins et~al.\ 1988\nocite{Huggins88})
\begin{equation}
H_{\mathrm{pe}} = k_{\mathrm{pe}}\,n_{\mathrm{H_2}},
\label{H_pe}
\end{equation}
where $k_{\mathrm{pe}}$ is a constant that depends on the properties of the
dust grains. Following Huggins et~al. \cite*{Huggins88} we adopt
$k_{\mathrm{pe}}$$=$1$\times$10$^{-26}$\,erg\,s$^{-1}$.
This heating mechanism is important in the cool, tenuous, outer parts of CSEs
around high mass loss rate stars.

The spatial extent of the molecular envelope is generally an
important input parameter, and the derived mass loss rate will depend on this.
The radial distribution of CO in the CSE was estimated using the modelling
presented in Mamon et~al.  \cite*{Mamon88}.  It includes
photodissociation, taking into account the effects of dust-, self- and
H$_{2}$-shielding, and chemical exchange reactions.
Mamon et~al.  \cite*{Mamon88} find that the radial abundance
distribution of CO with respect to H$_{2}$, $f(r)$, can be described by
\begin{equation}
f(r) = f_0 \, \exp \left[ - \ln 2 \left( \frac{r}{r_{\mathrm p}}
\right)^{\alpha}\, \right],
\end{equation}
where $f_0$ is the initial (photospheric) abundance,
$r_{\mathrm p}$ is the photodissociation radius (where the abundance
has dropped to $f_0/2$), and $\alpha$ is a parameter describing the
rate at which the abundance declines.  Both $r_{\mathrm p}$ and
$\alpha$ depend on the mass loss rate ($\dot{M}$), the expansion velocity
($v_{\mathrm e}$), and $f_0$.
In our modelling we use analytical fits to these results from
Stanek et al. \cite*{Stanek95}
\begin{eqnarray}
\label{rout}
r_{\mathrm p} & = & 5.4 \times 10^{16}
\left( \frac{\dot{M}}{10^{-6}} \right)^{0.65}
\left( \frac{v_{\mathrm e}}{15} \right)^{-0.55}
\left( \frac{f_0}{8\times 10^{-4}} \right)^{0.55} \nonumber \\
&& + \,7.5\times 10^{15} \left( \frac{v_{\mathrm e}}{15} \right)
\ {\mathrm {cm}}
\end{eqnarray}
and Kwan \& Webster \cite*{Kwan93}
\begin{equation}
\label{alpha}
\alpha = 2.79 \left( \frac{\dot{M}}{10^{-6}} \right)^{0.09}
               \left( \frac{15}{v_{\mathrm e}} \right)^{0.09},
\end{equation}
where the units of $\dot{M}$ and $v_{\mathrm e}$ are
${\mathrm M}_{\sun}\mathrm{yr}^{-1}$ and ${\mathrm {km\,s}}^{-1}$,
respectively.
In what follows we will assume
$f_0$=1$\times$10$^{-3}$.  This is an average of the $f_0$:s estimated
by Olofsson et~al.  \cite*{Olofsson93b} for this sample of optically
bright carbon stars. The scatter in the estimated $f_0$:s is
about 40\%. We note here that a somewhat more sophisticated
photodissociation model was developed by Doty \& Leung \cite*{Doty98}
in which they include scattering by dust.  For the CSE of
\object{CW~Leo} they derive a CO envelope size which is 30\% smaller than what
is obtained using the Mamon et al. model.  It is however not clear
by how much the envelope size changes for an object with a significantly lower
mass loss rate, and hence we use the results of Mamon et al. here.

In our models we include both a central source of radiation and the
cosmic microwave background radiation at 2.7\,K. The central radiation
emanates from the star, and was estimated from a fit to its spectral
energy distribution (SED), usually by assuming two blackbodies.
This method is described in Kerschbaum \cite*{Kerschbaum99}.
A fit to the SED
gives the two blackbody temperatures, $T_{*}$ and $T_{\rm d}$, and the
relative luminosities of the two blackbodies, $L_{\rm d}$/$L_{*}$.
Any dust present around the star will absorb parts of the stellar
light and re-emit it at longer wavelenghts. Thus, of the two blackbodies used,
one represents the stellar contribution and one represents the dust.
However, it should be noted that, e.g., the stellar temperature $T_{*}$
derived in this manner is generally lower than the true temperature of the
star (Kerschbaum 1999\nocite{Kerschbaum99}). In any case, the two
blackbodies used give a good description of the radiation that the
CSE is subjected to. The temperatures and luminosities used in the modelling
are presented in Table~\ref{coresult}. Only in the cases where
$L_{\rm d}$/$L_{*}$$>$0.1 the dust component was retained.
The inner boundary of the CSE, $r_{\rm i}$, was
set to reside outside the radius of the central blackbody(s), but
never lower than 1$\times$10$^{14}$\,cm ($\sim$3\,R$_{\sun}$), i.e., generally
beyond both the sonic point and dust condensation radius.

\begin{figure}
        \resizebox{\hsize}{!}{\includegraphics{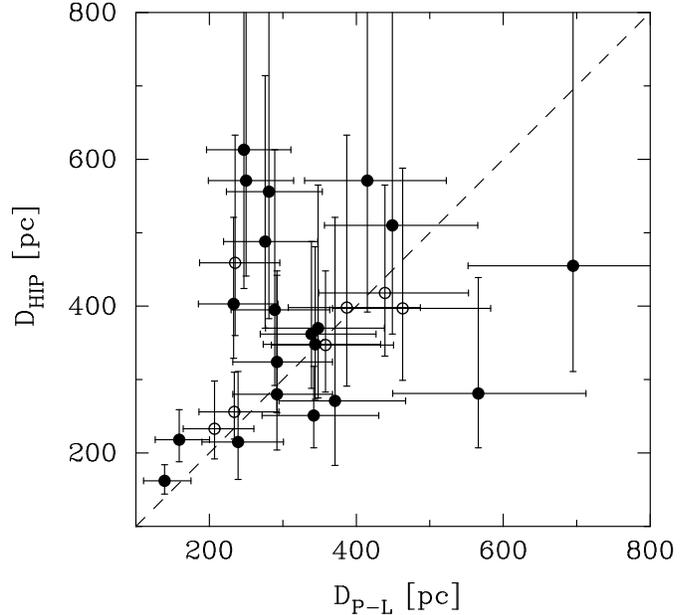}}
        \caption{Comparison between the distances derived from the
        period-luminosity relation ($D_{\mathrm{P-L}}$) and the Hipparcos
        parallaxes ($D_{\mathrm{HIP}}$). The irregular variables where the
        luminosity was assumed to equal 4000\,L$_{\sun}$ are indicated by
        open circles.  The dashed line shows the 1:1 correlation.}
        \label{hipp}
\end{figure}

The distances, presented in Table~\ref{coresult}, were estimated using
one of the following methods: the observed Hipparcos parallax, a
period-luminosity relation (Groenewegen \& Whitelock
1996\nocite{Groenewegen96}), or an assumed bolometric luminosity.  In
the two former cases the luminosities were estimated using apparent
bolometric magnitudes and the distances.  In the rare cases when both
the distance obtained from Hipparcos and the period-luminosity
relation result in extremely high or low luminosities, a value of
4000\,L$_{\sun}$ was adopted as the luminosity and the distance was
estimated using this value.  For the irregular variables with no
Hipparcos data we also assumed a luminosity of 4000\,L$_{\sun}$.
For a statistical study of a large sample of stars these distance
estimates are adequate, although the distance estimate for an
individual star has a large uncertainty.  There is no apparent
systematic difference between the distances derived from the
period-luminosity relation and the Hipparcos parallaxes,
Fig.~\ref{hipp}. The distances in Olofsson et~al.\
(1993a)\nocite{Olofsson93a} were estimated from an adopted absolute
$K$ magnitude.  The distances presented here are systematically lower
by, on the average, a factor of 1.4 than those used in Olofsson et~al.

This was a summary of the assumptions used in "the standard model".  In what
follows, tests will be made to see how sensitive the model is to the
various assumptions.

\begin{table*}
\caption[ ]{Input parameters and CO modelling results using the standard model 
(see text for details)}
\begin{tabular}{llcccccccccc}
\hline
\noalign{\smallskip}
 \multicolumn{1}{c}{Source} & 
 \multicolumn{1}{c}{Var. type} &
 \multicolumn{1}{c}{$P$}    & 
 \multicolumn{1}{c}{$D$}    & 
 \multicolumn{1}{c}{$L$}    & 
 \multicolumn{1}{c}{$T_*$}    & 
 \multicolumn{1}{c}{$T_{\mathrm d}^6$}    & 
 \multicolumn{1}{c}{$L_{\mathrm d}/L_*^6$}    & 
 \multicolumn{1}{c}{${\dot M}$} & 
 \multicolumn{1}{c}{$v_{\mathrm e}$}& 
 \multicolumn{1}{c}{$r_{\mathrm p}$} &
 \multicolumn{1}{c}{$h$}\\
  &
  &
 \multicolumn{1}{c}{[days]} & 
 \multicolumn{1}{c}{[pc]} & 
 \multicolumn{1}{c}{[L$_{\sun}$]} & 
 \multicolumn{1}{c}{[K]} & 
 \multicolumn{1}{c}{[K]} & &
 \multicolumn{1}{c}{$[\mathrm{M}_\odot\,\rm {yr}^{-1}]$} &
 \multicolumn{1}{c}{[km\,s$^{-1}$]} &
 \multicolumn{1}{c}{[cm]} &\\
\noalign{\smallskip}
\hline
\noalign{\smallskip}
\object{VX And}    & SRa & 369 & 560$^2$ &\phantom{0}5500 & 2000 &	             &	                        &  4.0$\times$10$^{-8}$ & 11.5  & 1.4$\times$10$^{16}$ & 0.2$^4$ \\
\object{HV Cas}    & M   & 527 & 970$^2$ &\phantom{0}7900 & 1800 &\phantom{0}900 &\phantom{0}0.54           &   9.0$\times$10$^{-7}$ & 18.5  & 6.1$\times$10$^{16}$ & 0.5$^4$ \\
\object{IRC+60041} &     &     & 670$^3$ &\phantom{0}4000 & 1300 &\phantom{0}500 &\phantom{0}0.13           &   3.0$\times$10$^{-6}$ & 28.0  & 1.0$\times$10$^{17}$ & 0.2$^4$ \\
\object{Z Psc}     & SRb & 144 & 320$^1$ &\phantom{0}3100 & 2600 &               &	                        & 2.5$\times$10$^{-8}$ &  \phantom{0}3.5  & 1.4$\times$10$^{16}$ & 0.2$^4$ \\
\object{R For}     & M   & 389 & 610$^2$ &\phantom{0}5800 & 1500 &\phantom{0}700 &\phantom{0}0.52           & 1.3$\times$10$^{-6}$ & 16.5  & 8.0$\times$10$^{16}$& 0.7\phantom{$^4$}  \\
\object{TW Hor}    & SRb & 158 & 400$^1$ &\phantom{0}6700 & 2600 &               &	                        &   9.0$\times$10$^{-8}$ &  \phantom{0}5.5  & 2.5$\times$10$^{16}$ & 0.5\phantom{$^4$}  \\
\object{V384 Per}  & M   & 535 & 560$^2$ &\phantom{0}8100 & 1900 &\phantom{0}900 &10\phantom{.00}           & 3.5$\times$10$^{-6}$ & 15.0  & 1.4$\times$10$^{17}$ & 0.4\phantom{$^4$}  \\
\object{V466 Per}  & SR  &     & 450$^3$ &\phantom{0}4000 & 2200 &               &	                        & 1.3$\times$10$^{-7}$ &  \phantom{0}9.0  & 2.6$\times$10$^{16}$ & 0.2$^4$ \\
\object{ST Cam}    & SRb & 300 & 330$^2$ &\phantom{0}4400 & 1600 &               &	                        &   9.0$\times$10$^{-8}$ &  \phantom{0}9.0  & 2.1$\times$10$^{16}$ & 0.2$^4$  \\
\object{TT Tau}    & SRb & 167 & 360$^2$ &\phantom{0}2400 & 2400 &               &	                        &   7.0$\times$10$^{-8}$ &  \phantom{0}5.0  & 2.2$\times$10$^{16}$ & 0.2$^4$  \\
\object{R Lep}     & M   & 467 & 250$^1$ &\phantom{0}4000 & 1900 &\phantom{0}800 &\phantom{0}0.14           &   7.0$\times$10$^{-7}$ & 18.0  & 5.3$\times$10$^{16}$ & 0.2\phantom{$^4$}  \\
\object{W Ori}     & SRb & 212 & 220$^1$ &\phantom{0}2600 & 2200 &               &	                        &   7.0$\times$10$^{-8}$ & 11.0  & 1.8$\times$10$^{16}$ & 0.3\phantom{$^4$}  \\
\object{S Aur}     & SR  & 590 & 820$^2$ &\phantom{0}8900 & 1600 &\phantom{0}700 &\phantom{0}0.32           & 2.2$\times$10$^{-6}$ & 25.5  & 8.9$\times$10$^{16}$ & 0.5$^4$  \\
\object{W Pic}     & Lb  &     & 490$^3$ &\phantom{0}4000 & 2000 &               &	                        &   3.0$\times$10$^{-7}$ & 16.0  & 3.5$\times$10$^{16}$ & 0.4\phantom{$^4$}  \\
\object{Y Tau}     & SRb & 242 & 310$^2$ &\phantom{0}3500 & 2400 &               &	                        &   4.0$\times$10$^{-7}$ & 11.0  & 4.5$\times$10$^{16}$ & 0.1 \\
\object{TU Gem}    & SRb & 230 & 330$^2$ &\phantom{0}3400 & 2400 &               &	                        & 2.5$\times$10$^{-7}$ & 11.5  & 3.4$\times$10$^{16}$ & 0.2$^4$ \\
\object{BL Ori}    & Lb  &     & 360$^3$ &\phantom{0}4000 & 2700 &               &	                        & 1.1$\times$10$^{-7}$ &  \phantom{0}9.0  & 2.4$\times$10$^{16}$ & 0.2$^4$  \\
\object{UU Aur}    & SRb & 234 & 260$^2$ &\phantom{0}6900 & 2500 &               &	                        & 3.5$\times$10$^{-7}$ & 11.0  & 4.2$\times$10$^{16}$ & 0.1\phantom{$^4$}	   \\
\object{NP Pup}    & Lb  &     & 420$^1$ &\phantom{0}3500 & 2700 &               &	                        & 6.5$\times$10$^{-8}$ &  \phantom{0}9.5  & 1.8$\times$10$^{16}$ & 0.2$^4$  \\
\object{CL Mon}    & M   & 497 & 770$^2$ &\phantom{0}7500 & 2000 &1100&\phantom{0}1.4\phantom{0}            & 2.2$\times$10$^{-6}$ & 25.0  & 8.9$\times$10$^{16}$ & 0.5$^4$  \\
\object{RY Mon}    & SRa & 456 & 456$^1$ &\phantom{0}3000 & 2000 &               &	                        & 3.5$\times$10$^{-7}$ & 11.0  & 4.2$\times$10$^{16}$ & 0.2$^4$  \\
\object{W CMa}     & Lb  &     & 450$^3$ &\phantom{0}4000 & 2500 &               &	                        & 3.0$\times$10$^{-7}$ & 10.5  & 3.9$\times$10$^{16}$ & 0.2$^4$  \\
\object{R Vol}     & M   & 454 & 730$^2$ &\phantom{0}6800 & 1300 &\phantom{0}800 &\phantom{0}0.7\phantom{0} & 1.8$\times$10$^{-6}$ & 18.0  & 9.0$\times$10$^{16}$ & 1.0\phantom{$^4$}  \\
\object{X Cnc}     & SRb & 195 & 280$^2$ &\phantom{0}2800 & 2200 &               &	                        & 1.0$\times$10$^{-7}$ &  \phantom{0}7.0  & 2.4$\times$10$^{16}$ & 0.1\phantom{$^4$}  \\
\object{CW Leo}    & M   & 630 & 120$^2$ &\phantom{0}9600 &      &\phantom{0}510 &	                        & 1.5$\times$10$^{-5}$ & 14.5  & 3.7$\times$10$^{17}$ & 1.0\phantom{$^4$}  \\
\object{Y Hya}     & SRb & 303 & 350$^1$ &\phantom{0}4200 & 2600 &               &	                        & 1.9$\times$10$^{-7}$ &  \phantom{0}9.0  & 3.2$\times$10$^{16}$ & 0.2$^4$  \\
\object{X Vel}     & SR  & 140 & 310$^2$ &\phantom{0}2800 & 2200 &               &	                        & 1.8$\times$10$^{-7}$ & 10.0  & 3.0$\times$10$^{16}$ & 0.2$^4$  \\
\object{SZ Car}    & SRb & 126 & 580$^3$ &\phantom{0}4000 & 2400 &               &	                        & 2.0$\times$10$^{-6}$ & 14.0  & 1.1$\times$10$^{17}$ & 0.2\phantom{$^4$}  \\
\object{RW LMi}    & SRa & 640 & 440$^2$ &\phantom{0}9700 & 1300 &\phantom{0}510 &\phantom{0}6.7\phantom{0} &   6.0$\times$10$^{-6}$ & 17.0  & 1.9$\times$10$^{17}$ & 1.4\phantom{$^4$}  \\
\object{XZ Vel}    &     &     & 530$^3$ &\phantom{0}4000 & 1900 &    	     &  	                &   6.0$\times$10$^{-7}$ & 14.0  & 5.2$\times$10$^{16}$ & 0.2$^4$ \\
\object{CZ Hya}    & M   & 442 & 960$^2$ &\phantom{0}6600 & 1800 &    	     &  	                &   9.0$\times$10$^{-7}$ & 12.0  & 7.0$\times$10$^{16}$ & 0.5$^4$ \\
\object{CCS 2792}  &     &     & 480$^3$ &\phantom{0}4000 & 2000 &    	     &  	                & 6.5$\times$10$^{-7}$ & 17.0  & 5.2$\times$10$^{16}$ & 0.2$^4$ \\
\object{U Hya}     & SRb & 450 & 160$^1$ &\phantom{0}2500 & 2400 &  	     &  	                & 1.4$\times$10$^{-7}$ &  \phantom{0}7.0  & 2.9$\times$10$^{16}$ & 0.2$^4$ \\
\object{VY UMa}    & Lb  &     & 330$^3$ &\phantom{0}4000 & 2700 &  	     &  	                &   7.0$\times$10$^{-8}$ &  \phantom{0}6.0  & 2.1$\times$10$^{16}$ & 0.2$^4$ \\
\object{SS Vir}    & SRa & 364 & 540$^2$ &\phantom{0}5400 & 1800 &  	     &  	                &   2.0$\times$10$^{-7}$ & 12.5  & 3.0$\times$10$^{16}$ & 0.2$^4$ \\
\object{Y CVn}     & SRb & 157 & 220$^1$ &\phantom{0}4400 & 2200 &  	     &  	                & 1.5$\times$10$^{-7}$ &  \phantom{0}8.5  & 2.9$\times$10$^{16}$ & 0.2$^5$  \\
\object{RY Dra}    & SRb:& 200 & 490$^1$ &10000           & 2500 &  	     &  	                &   3.0$\times$10$^{-7}$ & 10.0  & 4.0$\times$10$^{16}$ & 1.2$^5$  \\
\object{HD 121658} &     &     & 530$^3$ &\phantom{0}4000 & 2400 &  	     &  	                & 1.0$\times$10$^{-7}$ &  \phantom{0}6.5  & 2.5$\times$10$^{16}$ & 0.2$^4$  \\
\object{HD 124268} &     &     & 390$^3$ &\phantom{0}4000 & 2300 &               &	                        & 1.0$\times$10$^{-7}$ & 11.0  & 2.2$\times$10$^{16}$ & 0.2$^4$  \\
\object{X TrA}     & Lb  &     & 460$^3$ &\phantom{0}4000 & 2200 &               &	                        & 1.3$\times$10$^{-7}$ &  \phantom{0}8.0  & 2.7$\times$10$^{16}$ & 0.4\phantom{$^4$}  \\
\object{V CrB}     & M   & 358 & 630$^2$ &\phantom{0}5300 & 1900 &\phantom{0}740 &\phantom{0}0.15           &   6.0$\times$10$^{-7}$ &  \phantom{0}7.5  & 6.8$\times$10$^{16}$ & 0.2\phantom{$^4$}	\\
\object{TW Oph}    & SRb & 185 & 280$^1$ &\phantom{0}2700 & 2000 &               &	                        &  5.0$\times$10$^{-8}$ &  \phantom{0}7.5  & 1.6$\times$10$^{16}$ & 0.2$^4$  \\
\object{T Dra}     & M   & 422 & 610$^2$ &\phantom{0}6300 & 1600 &\phantom{0}650 &\phantom{0}0.28           & 1.2$\times$10$^{-6}$ & 13.5  & 8.0$\times$10$^{16}$ & 0.5$^4$   \\
\object{T Lyr}     & Lb  &     & 340$^3$ &\phantom{0}4000 & 2000 &               &	                        &  7.0$\times$10$^{-8}$ & 11.5  & 1.8$\times$10$^{16}$ & 0.5$^5$  \\
\object{V Aql}     & SRb & 353 & 370$^1$ &\phantom{0}6500 & 2100 &               &	                        &  3.0$\times$10$^{-7}$ &  \phantom{0}8.5  & 4.2$\times$10$^{16}$ & 0.2\phantom{$^4$}  \\
\object{V1942 Sgr} & Lb  &     & 430$^3$ &\phantom{0}4000 & 2600 &               &	                        & 1.6$\times$10$^{-7}$ & 10.0  & 2.8$\times$10$^{16}$ & 0.2$^4$  \\
\object{UX Dra}    & SRa:& 168 & 310$^3$ &\phantom{0}4000 & 2600 &               &	                        & 1.6$\times$10$^{-7}$ &  \phantom{0}4.0  & 4.0$\times$10$^{16}$ & 0.2$^4$  \\
\object{AQ Sgr}    & SRb & 200 & 420$^3$ &\phantom{0}4000 & 2700 &               &	                        & 2.5$\times$10$^{-7}$ & 10.0  & 3.6$\times$10$^{16}$ & 0.2$^4$  \\
\object{RT Cap}    & SRb & 393 & 450$^2$ &\phantom{0}5900 & 2200 &               &	                        & 1.0$\times$10$^{-7}$ &  \phantom{0}8.0  & 2.3$\times$10$^{16}$ & 0.2$^4$  \\
\object{U Cyg}     & M   & 463 & 710$^2$ &\phantom{0}6900 & 2200 &               &	                        & 9.0$\times$10$^{-7}$ & 13.0  & 6.8$\times$10$^{16}$ & 0.5$^4$\\
\object{V Cyg}     & M   & 421 & 370$^2$ &\phantom{0}6200 & 1500 &\phantom{0}860 &\phantom{0}0.84           & 1.2$\times$10$^{-6}$ & 11.5  & 8.5$\times$10$^{16}$ & 1.2\phantom{$^4$}  \\
\object{RV Aqr}    & M   & 454 & 670$^2$ &\phantom{0}6800 & 1300 &\phantom{0}620 &\phantom{0}0.46           & 2.5$\times$10$^{-6}$ & 16.0  & 1.1$\times$10$^{17}$ & 0.5\phantom{$^4$}  \\
\object{T Ind}     & SRb & 320 & 570$^1$ &\phantom{0}8800 & 2600 &               &	                        &  9.0$\times$10$^{-8}$ &  \phantom{0}6.0  & 2.4$\times$10$^{16}$ & 0.5$^4$ \\
\object{Y Pav}     & SRb & 233 & 360$^2$ &\phantom{0}4100 & 2400 &               &	                        & 1.6$\times$10$^{-7}$ &  \phantom{0}8.0  & 3.0$\times$10$^{16}$ & 0.2$^4$ \\
\object{V1426 Cyg} & M   & 470 & 780$^2$ &\phantom{0}7100 & 1200 &\phantom{0}580 &\phantom{0}0.5\phantom{0}& 1.0$\times$10$^{-5}$ & 14.0  & 2.9$\times$10$^{17}$ & 0.3\phantom{$^4$}  \\
\object{S Cep}     & M   & 487 & 340$^2$ &\phantom{0}7300 & 1400 &               &	                        & 1.5$\times$10$^{-6}$ & 22.0  & 7.5$\times$10$^{16}$ & 0.4\phantom{$^4$}  \\
\noalign{\smallskip}
\hline
\end{tabular}
\label{coresult}
\end{table*}

\addtocounter{table}{-1}
\begin{table*}
\caption[ ]{continued.}
\begin{tabular}{llcccccccccc}
\hline
\noalign{\smallskip}
 \multicolumn{1}{c}{Source} & 
 \multicolumn{1}{c}{Var. type} &
 \multicolumn{1}{c}{$P$}    & 
 \multicolumn{1}{c}{$D$}    & 
 \multicolumn{1}{c}{$L$}    & 
 \multicolumn{1}{c}{$T_*$}    & 
 \multicolumn{1}{c}{$T_{\mathrm d}^6$}    & 
 \multicolumn{1}{c}{$L_{\mathrm d}/L_*^6$}    & 
 \multicolumn{1}{c}{${\dot M}$} & 
 \multicolumn{1}{c}{$v_{\mathrm e}$}& 
 \multicolumn{1}{c}{$r_{\mathrm p}$} &
 \multicolumn{1}{c}{$h$}\\
  &
  &
 \multicolumn{1}{c}{[days]} & 
 \multicolumn{1}{c}{[pc]} & 
 \multicolumn{1}{c}{[L$_{\sun}$]} & 
 \multicolumn{1}{c}{[K]} & 
 \multicolumn{1}{c}{[K]} & &
 \multicolumn{1}{c}{$[\mathrm{M}_\odot\,\rm {yr}^{-1}]$} &
 \multicolumn{1}{c}{[km\,s$^{-1}$]} &
 \multicolumn{1}{c}{[cm]} &\\
\noalign{\smallskip}
\hline
\noalign{\smallskip}
\object{V460 Cyg}  & SRb & 180 & 230$^2$ &\phantom{0}2600 & 2800 &               &	              	        & 1.8$\times$10$^{-7}$ & 10.0  & 3.0$\times$10$^{16}$ & 0.2$^4$ \\
\object{RV Cyg}    & SRb & 263 & 310$^2$ &\phantom{0}3400 & 2100 &               &	              	        & 4.5$\times$10$^{-7}$ & 13.5  & 4.5$\times$10$^{16}$ & 0.2$^4$ \\
\object{PQ Cep}    & M   &     & 390$^2$ &\phantom{0}4000 & 1700 &\phantom{0}840 &\phantom{0}0.13           & 1.4$\times$10$^{-6}$ & 19.5  & 7.6$\times$10$^{16}$ & 0.2$^4$  \\
\object{LP And}    & M   & 620 & 630$^2$ &\phantom{0}9400 & 1100 &\phantom{0}610 &\phantom{0}6.6\phantom{0} & 1.5$\times$10$^{-5}$ & 14.0  & 4.0$\times$10$^{17}$& 0.7\phantom{$^4$}  \\
\object{WZ Cas}    & SRb & 186 & 290$^2$ &\phantom{0}2700 & 2500 &               &                          & 6.5$\times$10$^{-9}$ & \phantom{0}2.5  & 7.4$\times$10$^{15}$ & 0.2$^4$  \\
\noalign{\smallskip}
\hline
\noalign{\smallskip}
\noalign{
  $^1$ Hipparcos; $^2$ Period-luminosity relation 
  (Groenewegen \& Whitelock 1996\nocite{Groenewegen96}); 
  $^3$ $L$$=$4000\,L$_{\sun}$ assumed;
  $^4$ Assumed $h$-parameter (see text for details);
  $^5$ J-type star (only including $^{12}$CO cooling);
  $^6$ In the cases when $L_{\mathrm d}/L_*$$<$0.1 the dust blackbody component 
       was not retained and a single blackbody representing the stellar component 
       adequately describes the SED. In the case of \object{CW~Leo} a single 
       dust blackbody is used to represent the SED.}
\end{tabular}
\end{table*}

\begin{table*}
\caption[ ]{CO modelling results compared to single dish observations.}
\begin{flushleft}
\begin{tabular}{lllccc|lllccc}
\hline
 \multicolumn{1}{c}{Source} & 
 \multicolumn{1}{c}{Tel.} & 
 \multicolumn{1}{c}{Trans.}& 
 \multicolumn{1}{c}{$I_{\mathrm{obs}}$} &
 \multicolumn{1}{c}{$I_{\mathrm{mod}}$} &
 \multicolumn{1}{c|}{Ref.} &
 \multicolumn{1}{c}{Source} & 
 \multicolumn{1}{c}{Tel.} & 
 \multicolumn{1}{c}{Trans.}& 
 \multicolumn{1}{c}{$I_{\mathrm{obs}}$} &
 \multicolumn{1}{c}{$I_{\mathrm{mod}}$} &
 \multicolumn{1}{c}{Ref.}\\
  &
  &
  &
 \multicolumn{1}{c}{[K\,km\,s$^{-1}$]} &
 \multicolumn{1}{c}{[K\,km\,s$^{-1}$]} & 
 \multicolumn{1}{c|}{}&
  &
  &
  &
 \multicolumn{1}{c}{[K\,km\,s$^{-1}$]} &
 \multicolumn{1}{c}{[K\,km\,s$^{-1}$]} & \\
\hline
%
%
\object{VX And}    & OSO   & 1$-$0 & \phantom{000}$<$0.8 & \phantom{000}0.1 & 2&\object{UU Aur}    & OSO   & 1$-$0 & \phantom{000$<$}7.9 & \phantom{000}8.2 & 2\\
                   & IRAM  & 1$-$0 & \phantom{000}$<$1.4 & \phantom{000}0.2 & 2&		   & IRAM  & 1$-$0 & \phantom{00$<$}18.8 & \phantom{00}16.9 & 2\\
                   & IRAM  & 2$-$1 & \phantom{000$<$}2.4 & \phantom{000}2.4 & 2&		   & JCMT  & 2$-$1 & \phantom{00$<$}16.8 & \phantom{00}14.8 & 4\\
\object{HV Cas}    & OSO   & 1$-$0 & \phantom{000}$<$2.1 & \phantom{000}1.9 & 2&                   & IRAM  & 2$-$1 & \phantom{00$<$}39.0 & \phantom{00}46.2 & 2\\
                   & IRAM  & 1$-$0 & \phantom{000$<$}5.6 & \phantom{000}4.1 & 2&\object{NP Pup}    & SEST  & 1$-$0 & \phantom{000}$<$0.5 & \phantom{000}0.3 & 2 \\
                   & IRAM  & 2$-$1 & \phantom{00$<$}11.9 & \phantom{00}14.5 & 2& 		   & SEST  & 2$-$1 & \phantom{000$<$}1.2 & \phantom{000}1.2 & 2 \\
\object{IRC+60041} & IRAM  & 1$-$0 & \phantom{00$<$}IS\phantom{.0}& \phantom{000}4.7 &2&\object{CL Mon}	& SEST  & 1$-$0 &\phantom{000$<$}3.2 & \phantom{000}3.3 & 2\\
                   & IRAM  & 2$-$1 & \phantom{00$<$}17.3 & \phantom{00}17.3 & 2& 	           & SEST  & 2$-$1 & \phantom{00$<$}10.4 & \phantom{000}9.5 & 2 \\
\object{Z Psc}     & OSO   & 1$-$0 & \phantom{000$<$}1.0 & \phantom{000}0.7 & 2& 		   & IRAM  & 2$-$1 & \phantom{00$<$}34.3 & \phantom{00}38.1 & 2 \\
                   & IRAM  & 1$-$0 & \phantom{000$<$}2.1 & \phantom{000}1.6 & 2&\object{RY Mon}    & IRAM  & 1$-$0 & \phantom{000$<$}6.8 & \phantom{000}5.4 & 2 \\
                   & IRAM  & 2$-$1 & \phantom{000$<$}4.6 & \phantom{000}6.0 & 2& 		   & IRAM  & 2$-$1 & \phantom{00$<$}11.3 & \phantom{00}14.5 & 2 \\
\object{R For}     & SEST  & 1$-$0 & \phantom{000$<$}4.8 & \phantom{000}4.9 & 2&\object{W CMa}     & SEST  & 1$-$0 & \phantom{000$<$}1.3 & \phantom{000}1.3 & 2 \\
                   & SEST  & 2$-$1 & \phantom{00$<$}11.0 & \phantom{00}14.5 & 2& 		   & SEST  & 2$-$1 & \phantom{000$<$}4.6 & \phantom{000}3.9 & 2 \\
		   & JCMT  & 2$-$1 & \phantom{00$<$}21.0 & \phantom{00}17.3 & 4& 		   & IRAM  & 2$-$1 & \phantom{00$<$}11.7 & \phantom{000}16.0 & 2 \\
		   & SEST  & 3$-$2 & \phantom{00$<$}18.0 & \phantom{00}19.9 & 1&\object{R Vol}     & SEST  & 1$-$0 & \phantom{000$<$}4.9 & \phantom{000}5.1 & 2 \\
		   & JCMT  & 3$-$2 & \phantom{00$<$}27.4 & \phantom{00}25.6 & 4& 		   & SEST  & 2$-$1 & \phantom{00$<$}15.3 & \phantom{000}14.7 & 2 \\
\object{TW Hor}    & SEST  & 1$-$0 & \phantom{000$<$}1.0 & \phantom{000}0.9 & 2& 		   & SEST  & 3$-$2 & \phantom{00$<$}21.0 & \phantom{000}22.2 & 1 \\
                   & SEST  & 2$-$1 & \phantom{000$<$}3.3 & \phantom{000}3.8 & 2&\object{X Cnc}	   & NRAO  & 1$-$0 & \phantom{000$<$}0.7 & \phantom{000}0.7 & 1\\
		   & SEST  & 3$-$2 & \phantom{000$<$}7.2 & \phantom{000}6.1 & 1& 		   & SEST  & 1$-$0 & \phantom{000$<$}1.5 & \phantom{000}1.1 & 2\\
\object{V384 Per}  & NRAO  & 1$-$0 & \phantom{000$<$}7.8 & \phantom{00}10.1 & 1& 		   & OSO   & 1$-$0 & \phantom{000$<$}1.7 & \phantom{000}2.0 & 2\\
                   & OSO   & 1$-$0 & \phantom{00$<$}25.5 & \phantom{00}25.1 & 2& 		   & SEST  & 2$-$1 & \phantom{000$<$}3.2 & \phantom{000}2.9 & 2\\
                   & IRAM  & 1$-$0 & \phantom{00$<$}52.7 & \phantom{00}48.4 & 2& 		   & IRAM  & 2$-$1 & \phantom{00$<$}11.1 & \phantom{00}11.8 & 2\\
		   & JCMT  & 2$-$1 & \phantom{00$<$}37.3 & \phantom{00}35.2 & 4&\object{CW Leo}    & NRAO  & 1$-$0 & \phantom{0$<$}170.8 & \phantom{0}257.1 & 1\\
		   & IRAM  & 2$-$1 & \phantom{00$<$}83.6 & \phantom{0}102.0 & 2&                   & SEST  & 1$-$0 & \phantom{0$<$}288.1 & \phantom{0}305.3 & 2\\
		   & JCMT  & 3$-$2 & \phantom{00$<$}44.5 & \phantom{00}42.5 & 4&		   & OSO   & 1$-$0 & \phantom{0$<$}422.0 & \phantom{0}391.7 & 2\\
\object{V466 Per}  & IRAM  & 1$-$0 & \phantom{000$<$}3.6 & \phantom{000}2.5 & 2&		   & SEST  & 2$-$1 & \phantom{0$<$}487.3 & \phantom{0}580.5 & 2\\
                   & IRAM  & 2$-$1 & \phantom{000$<$}6.2 & \phantom{000}8.6 & 2&		   & JCMT  & 2$-$1 & \phantom{0$<$}731.8 & \phantom{0}632.2 & 4\\
\object{ST Cam}    & OSO   & 1$-$0 & \phantom{000$<$}2.0 & \phantom{000}2.0 & 2&		   & IRAM  & 2$-$1 & \phantom{$<$}1057.7 & 	     1116.1 & 2\\
                   & JCMT  & 2$-$1 & \phantom{000$<$}3.4 & \phantom{000}4.4 & 4&                   & SEST  & 3$-$2 & \phantom{0$<$}854.8 & \phantom{0}774.0 & 1\\
		   & IRAM  & 2$-$1 & \phantom{00$<$}15.2 & \phantom{00}14.8 & 2&		   & JCMT  & 3$-$2 & \phantom{$<$}1066.3 & \phantom{0}883.7 & 4\\
\object{TT Tau}    & IRAM  & 1$-$0 & \phantom{000$<$}2.6 & \phantom{000}2.7 & 2&		   & JCMT  & 4$-$3 & \phantom{$<$}1227.8 & 	     1052.0 & 4\\
                   & IRAM  & 2$-$1 & \phantom{000$<$}4.0 & \phantom{000}8.1 & 2&\object{Y Hya}     & SEST  & 1$-$0 & \phantom{000$<$}1.6 & \phantom{000}1.6 & 2\\
\object{R Lep}     & SEST  & 1$-$0 & \phantom{000$<$}6.2 & \phantom{000}6.7 & 2&                   & IRAM  & 2$-$1 & \phantom{00$<$}18.7 & \phantom{00}19.4 & 2\\
                   & IRAM  & 1$-$0 & \phantom{00$<$}31.7 & \phantom{00}24.3 & 2&\object{X Vel}     & SEST  & 1$-$0 & \phantom{000$<$}1.4 & \phantom{000}1.5 & 2\\
		   & SEST  & 2$-$1 & \phantom{00$<$}18.1 & \phantom{00}17.4 & 2&                   & SEST  & 2$-$1 & \phantom{000$<$}4.3 & \phantom{000}4.4 & 2\\
                   & IRAM  & 2$-$1 & \phantom{00$<$}56.4 & \phantom{00}64.2 & 2&\object{SZ Car}    & SEST  & 1$-$0 & \phantom{000$<$}2.8 & \phantom{000}2.7 & 2\\
		   & SEST  & 3$-$2 & \phantom{00$<$}12.5 & \phantom{00}21.6 & 2&                   & SEST  & 2$-$1 & \phantom{000$<$}7.8 & \phantom{000}5.9 & 2\\
		   & JCMT  & 3$-$2 & \phantom{00$<$}28.1 & \phantom{00}27.9 & 4&                   & SEST  & 3$-$2 & \phantom{000$<$}5.9 & \phantom{000}6.2 & 1\\
\object{W Ori}     & SEST  & 1$-$0 & \phantom{000$<$}1.2 & \phantom{000}1.3 & 2&\object{RW LMi}    & NRAO  & 1$-$0 & \phantom{00$<$}36.4 & \phantom{00}37.0 & 1\\
                   & OSO   & 1$-$0 & \phantom{000$<$}2.7 & \phantom{000}2.4 & 2&                   & SEST  & 1$-$0 & \phantom{00$<$}54.6 & \phantom{00}51.9 & 2\\
		   & IRAM  & 1$-$0 & \phantom{000$<$}5.0 & \phantom{000}5.1 & 2&		   & OSO   & 1$-$0 & \phantom{00$<$}87.0 & \phantom{00}83.8 & 2\\
		   & SEST  & 2$-$1 & \phantom{000$<$}4.9 & \phantom{000}4.3 & 2&                   & SEST  & 2$-$1 & \phantom{0$<$}105.4 & \phantom{0}128.4 & 2\\
                   & IRAM  & 2$-$1 & \phantom{00$<$}19.0 & \phantom{00}16.9 & 2&		   & JCMT  & 2$-$1 & \phantom{0$<$}163.2 & \phantom{0}147.4 & 4\\
		   & SEST  & 3$-$2 & \phantom{000$<$}4.8 & \phantom{000}5.4 & 1&		   & JCMT  & 3$-$2 & \phantom{0$<$}245.3 & \phantom{0}208.6 & 4\\
\object{S Aur}     & OSO   & 1$-$0 & \phantom{000$<$}6.7 & \phantom{000}5.5 & 2&                   & JCMT  & 4$-$3 & \phantom{0$<$}243.1 & \phantom{0}224.3 & 4\\
		   & IRAM  & 1$-$0 & \phantom{00$<$}10.9 & \phantom{00}11.9 & 2&\object{XZ Vel}    & SEST  & 1$-$0 & \phantom{000$<$}1.8 & \phantom{000}1.4 & 2\\
		   & IRAM  & 2$-$1 & \phantom{00$<$}15.8 & \phantom{00}36.8 & 2&                   & SEST  & 2$-$1 & \phantom{000$<$}3.5 & \phantom{000}4.1 & 2\\
\object{W Pic}     & SEST  & 1$-$0 & \phantom{000$<$}0.9 & \phantom{000}0.9 & 2&\object{CZ Hya}    & SEST  & 1$-$0 & \phantom{000$<$}1.5 & \phantom{000}1.5 & 2\\
                   & SEST  & 2$-$1 & \phantom{000$<$}2.6 & \phantom{000}3.1 & 2&                   & SEST  & 2$-$1 & \phantom{000$<$}4.2 & \phantom{000}4.5 & 2\\
                   & SEST  & 3$-$2 & \phantom{000$<$}6.1 & \phantom{000}4.5 & 1&\object{CCS 2792}  & SEST  & 1$-$0 & \phantom{000$<$}1.9 & \phantom{000}1.8 & 2\\
\object{Y Tau}     & SEST  & 1$-$0 & \phantom{000$<$}3.2 & \phantom{000}3.0 & 2&                   & SEST  & 2$-$1 & \phantom{000$<$}4.1 & \phantom{000}4.7 & 2\\
                   & OSO   & 1$-$0 & \phantom{000$<$}6.3 & \phantom{000}5.3 & 2&\object{U Hya}     & SEST  & 1$-$0 & \phantom{000$<$}5.4 & \phantom{000}5.4 & 2\\
		   & IRAM  & 1$-$0 & \phantom{00$<$}15.5 & \phantom{00}11.0 & 2&                   & SEST  & 2$-$1 & \phantom{00$<$}13.8 & \phantom{00}14.1 & 1\\
		   & SEST  & 2$-$1 & \phantom{000$<$}5.8 & \phantom{000}6.1 & 2&		   & JCMT  & 2$-$1 & \phantom{00$<$}20.2 & \phantom{00}16.6 & 4\\
                   & IRAM  & 2$-$1 & \phantom{00$<$}19.4 & \phantom{00}24.1 & 2&		   & IRAM  & 2$-$1 & \phantom{00$<$}48.8 & \phantom{00}48.7 & 4\\
\object{TU Gem}    & OSO   & 1$-$0 & \phantom{000$<$}3.4 & \phantom{000}3.4 & 2&\object{VY UMa}    & OSO   & 1$-$0 & \phantom{000$<$}1.4 & \phantom{000}1.4 & 2\\
		   & IRAM  & 1$-$0 & \phantom{000$<$}6.5 & \phantom{00}21.1 & 2&		   & IRAM  & 2$-$1 & \phantom{000$<$}4.3 & \phantom{00}11.0 & 2\\
\object{BL Ori}    & OSO   & 1$-$0 & \phantom{00$<$}IS\phantom{.0}& \phantom{000}1.2 & 2& \object{SS Vir}    & OSO   & 1$-$0& \phantom{000}$<$1.2 & \phantom{000}1.1 & 2 \\
		   & IRAM  & 1$-$0 & \phantom{00$<$}IS\phantom{.0}& \phantom{000}2.6 & 2&		     & IRAM  & 1$-$0& \phantom{000$<$}7.9 & \phantom{000}8.5 & 2\\
		   & IRAM  & 2$-$1 & \phantom{000$<$}9.8 & \phantom{000}9.7 & 2& \\
\hline										 
\end{tabular}
\end{flushleft}
\end{table*}

\addtocounter{table}{-1}
\begin{table*}
\caption[ ]{continued.}
\begin{flushleft}
\begin{tabular}{lllccc|lllccc}
\hline
 \multicolumn{1}{c}{Source} & 
 \multicolumn{1}{c}{Tel.} & 
 \multicolumn{1}{c}{Trans.}& 
 \multicolumn{1}{c}{$I_{\mathrm{obs}}$} &
 \multicolumn{1}{c}{$I_{\mathrm{mod}}$} &
 \multicolumn{1}{c|}{Ref.} &
 \multicolumn{1}{c}{Source} & 
 \multicolumn{1}{c}{Tel.} & 
 \multicolumn{1}{c}{Trans.}& 
 \multicolumn{1}{c}{$I_{\mathrm{obs}}$} &
 \multicolumn{1}{c}{$I_{\mathrm{mod}}$} &
 \multicolumn{1}{c}{Ref.}\\
  &
  &
  &
 \multicolumn{1}{c}{[K\,km\,s$^{-1}$]} &
 \multicolumn{1}{c}{[K\,km\,s$^{-1}$]} & 
 \multicolumn{1}{c|}{}&
  &
  &
  &
 \multicolumn{1}{c}{[K\,km\,s$^{-1}$]} &
 \multicolumn{1}{c}{[K\,km\,s$^{-1}$]} & \\
\hline
%
\object{Y CVn}    & OSO  & 1$-$0 & \phantom{000$<$}4.5 & \phantom{000}5.4 & 2&\object{V Cyg }	& OSO  & 1$-$0 & \phantom{00$<$}27.5 & \phantom{00}27.2 & 2\\
                  & IRAM & 1$-$0 & \phantom{00$<$}10.3 & \phantom{00}11.2 & 2&  		& NRAO & 2$-$1 & \phantom{00$<$}50.1 & \phantom{00}36.5 & 1\\
		  & JCMT & 2$-$1 & \phantom{00$<$}11.9 & \phantom{00}10.9 & 4&  		& JCMT & 2$-$1 & \phantom{00$<$}35.9 & \phantom{00}56.2 & 4\\
		  & IRAM & 2$-$1 & \phantom{00$<$}22.7 & \phantom{00}34.2 & 2&  		& JCMT & 3$-$2 & \phantom{00$<$}88.9 & \phantom{00}86.1 & 4\\
                  & JCMT & 3$-$2 & \phantom{00$<$}20.9 & \phantom{00}15.6 & 4&  		& JCMT & 4$-$3 & \phantom{0$<$}123.4 & \phantom{00}95.2 & 4\\
\object{RY Dra}   & OSO  & 1$-$0 & \phantom{000$<$}2.4 & \phantom{000}3.0 & 2&\object{RV Aqr}	& SEST & 1$-$0 & \phantom{000$<$}7.5 & \phantom{000}6.9 & 2\\
                  & IRAM & 2$-$1 & \phantom{00$<$}21.6 & \phantom{00}28.3 & 2&  		& SEST & 2$-$1 & \phantom{00$<$}18.1 & \phantom{00}17.2 & 2\\
                  & JCMT & 3$-$2 & \phantom{00$<$}18.9 & \phantom{00}15.8 & 4&  		& SEST & 3$-$2 & \phantom{00$<$}18.6 & \phantom{00}20.5 & 1\\
\object{HD 121658}& SEST & 1$-$0 & \phantom{000$<$}1.5 & \phantom{000}1.4 & 2&\object{T Ind}	& SEST & 1$-$0 & \phantom{000$<$}0.5 & \phantom{000}0.4 & 2\\
\object{HD 124268}& SEST & 1$-$0 & \phantom{000}$<$0.7 & \phantom{000}0.6 & 2&  		& SEST & 2$-$1 & \phantom{000$<$}1.5 & \phantom{000}1.9 & 2\\
                  & SEST & 2$-$1 & \phantom{000$<$}2.1 & \phantom{000}1.9 & 2&\object{Y Pav}	& SEST & 1$-$0 & \phantom{000$<$}1.6 & \phantom{000}1.4 & 2\\
\object{X TrA}    & SEST & 1$-$0 & \phantom{000$<$}2.5 & \phantom{000}2.7 & 2&  		& SEST & 2$-$1 & \phantom{000$<$}3.8 & \phantom{000}4.3 & 2\\
                  & SEST & 2$-$1 & \phantom{00$<$}11.8 & \phantom{00}10.2 & 2&\object{V1426 Cyg}& OSO  & 1$-$0 & \phantom{00$<$}17.3 & \phantom{00}17.0	& 2\\
		  & SEST & 3$-$2 & \phantom{00$<$}15.3 & \phantom{00}15.1 & 1&  		& SEST & 2$-$1 & \phantom{000$<$}8.7 & \phantom{00}14.3	& 2\\
\object{V CrB}    & OSO  & 1$-$0 & \phantom{000$<$}2.7 & \phantom{000}3.9 & 2&  		& SEST & 3$-$2 & \phantom{00$<$}17.8 & \phantom{00}17.1	& 1\\
                  & IRAM & 1$-$0 & \phantom{00$<$}11.4 & \phantom{000}8.4 & 3&\object{S Cep}	& OSO  & 1$-$0 & \phantom{00$<$}16.3 & \phantom{00}19.4 & 2\\
		  & IRAM & 2$-$1 & \phantom{00$<$}18.4 & \phantom{00}18.8 & 2&  		& IRAM & 1$-$0 & \phantom{00$<$}48.5 & \phantom{00}38.1 & 3\\
\object{TW Oph}   & SEST & 1$-$0 & \phantom{000$<$}1.1 & \phantom{000}0.9 & 2&  		& JCMT & 2$-$1 & \phantom{00$<$}49.3 & \phantom{00}37.3 & 4\\
                  & IRAM & 2$-$1 & \phantom{000$<$}8.7 & \phantom{00}10.2 & 2&  		& IRAM & 2$-$1 & \phantom{00$<$}76.6 & \phantom{0}109.6 & 2\\
\object{T Dra}    & OSO  & 1$-$0 & \phantom{000$<$}8.8 & \phantom{000}8.4 & 2&  		& JCMT & 3$-$2 & \phantom{00$<$}55.7 & \phantom{00}55.4 & 4\\
                  & IRAM & 1$-$0 & \phantom{00$<$}28.6 & \phantom{00}17.7 & 3&  		& JCMT & 4$-$3 & \phantom{00$<$}65.7 & \phantom{00}60.0 & 4\\
		  & IRAM & 2$-$1 & \phantom{00$<$}43.7 & \phantom{00}48.7 & 2&\object{V460 Cyg} & OSO  & 1$-$0 & \phantom{000$<$}4.7 & \phantom{000}4.8	& 2\\
\object{T Lyr}    & OSO  & 1$-$0 & \phantom{000$<$}0.7 & \phantom{000}1.0 & 2&  		& IRAM & 2$-$1 & \phantom{000$<$}8.5 & \phantom{00}28.4 & 2\\
                  & IRAM & 2$-$1 & \phantom{000$<$}5.6 & \phantom{000}9.6 & 2&\object{RV Cyg}	& OSO  & 1$-$0 & \phantom{000$<$}4.8 & \phantom{000}6.1	& 2\\
                  & JCMT & 3$-$2 & \phantom{000$<$}6.4 & \phantom{000}4.3 & 4&  		& IRAM & 1$-$0 & \phantom{00$<$}16.7 & \phantom{00}12.9	& 2\\
\object{V Aql}    & SEST & 1$-$0 & \phantom{000$<$}2.8 & \phantom{000}2.6 & 2&  		& IRAM & 2$-$1 & \phantom{00$<$}14.4 & \phantom{00}33.5 & 2\\
                  & OSO  & 1$-$0 & \phantom{000$<$}3.2 & \phantom{000}4.8 & 2&\object{PQ Cep}   & OSO  & 1$-$0 & \phantom{000$<$}6.4 & \phantom{000}6.7 & 2\\
		  & SEST & 2$-$1 & \phantom{000$<$}8.1 & \phantom{000}7.0 & 2&  		& NRAO & 2$-$1 & \phantom{000$<$}8.7 & \phantom{000}7.4 & 1\\
		  & JCMT & 2$-$1 & \phantom{000$<$}9.0 & \phantom{000}8.3 & 4&\object{LP And}	& OSO  & 1$-$0 & \phantom{00$<$}57.0 & \phantom{00}58.8 & 1\\
		  & JCMT & 3$-$2 & \phantom{00$<$}11.2 & \phantom{00}10.6 & 4&  		& IRAM & 1$-$0 & \phantom{00$<$}92.7 & \phantom{00}97.4 & 3\\
\object{V1942 Sgr}& SEST & 1$-$0 & \phantom{000$<$}0.8 & \phantom{000}0.8 & 2&  		& NRAO & 2$-$1 & \phantom{00$<$}60.7 & \phantom{00}46.5 & 1\\
\object{UX Dra}   & OSO  & 1$-$0 & \phantom{000$<$}2.1 & \phantom{000}1.7 & 2&  		& JCMT & 2$-$1 & \phantom{00$<$}90.6 & \phantom{00}68.5 & 4\\
                  & IRAM & 2$-$1 & \phantom{000$<$}6.8 & \phantom{000}9.0 & 2&  		& IRAM & 2$-$1 & \phantom{0$<$}155.8 & \phantom{0}166.8 & 3\\
\object{AQ Sgr}   & SEST & 1$-$0 & \phantom{000$<$}1.4 & \phantom{000}1.3 & 2&  		& JCMT & 3$-$2 & \phantom{00$<$}88.0 & \phantom{00}79.1 & 4\\
                  & SEST & 2$-$1 & \phantom{000$<$}4.1 & \phantom{000}4.0 & 2&  		& JCMT & 4$-$3 & \phantom{00$<$}73.3 & \phantom{00}78.6 & 4\\
\object{RT Cap}   & SEST & 1$-$0 & \phantom{000$<$}0.6 & \phantom{000}0.6 & 2&\object{WZ Cas}	& OSO  & 1$-$0 & \phantom{000}$<$1.1 & \phantom{000}0.1 & 2\\
                  & SEST & 2$-$1 & \phantom{000$<$}2.0 & \phantom{000}2.1 & 2&  		& IRAM & 1$-$0 & \phantom{000}$<$2.6 & \phantom{000}0.3 & 2\\
\object{U Cyg}    & OSO  & 1$-$0 & \phantom{000$<$}4.9 & \phantom{000}4.8 & 2&  		& IRAM & 2$-$1 & \phantom{000$<$}2.6 & \phantom{000}2.8 & 2\\ 
                  & IRAM & 2$-$1 & \phantom{000$<$}9.4 & \phantom{00}31.7 & 2\\

\hline
\noalign{\smallskip}
\noalign{1.\ This paper; 2.\ Olofsson et al.\ \cite*{Olofsson93a}; 
3. Neri et al. \cite*{Neri}; 4.\ JCMT public archive.}
\end{tabular}
\end{flushleft}
\end{table*}

\begin{figure}
        \resizebox{\hsize}{!}{\includegraphics{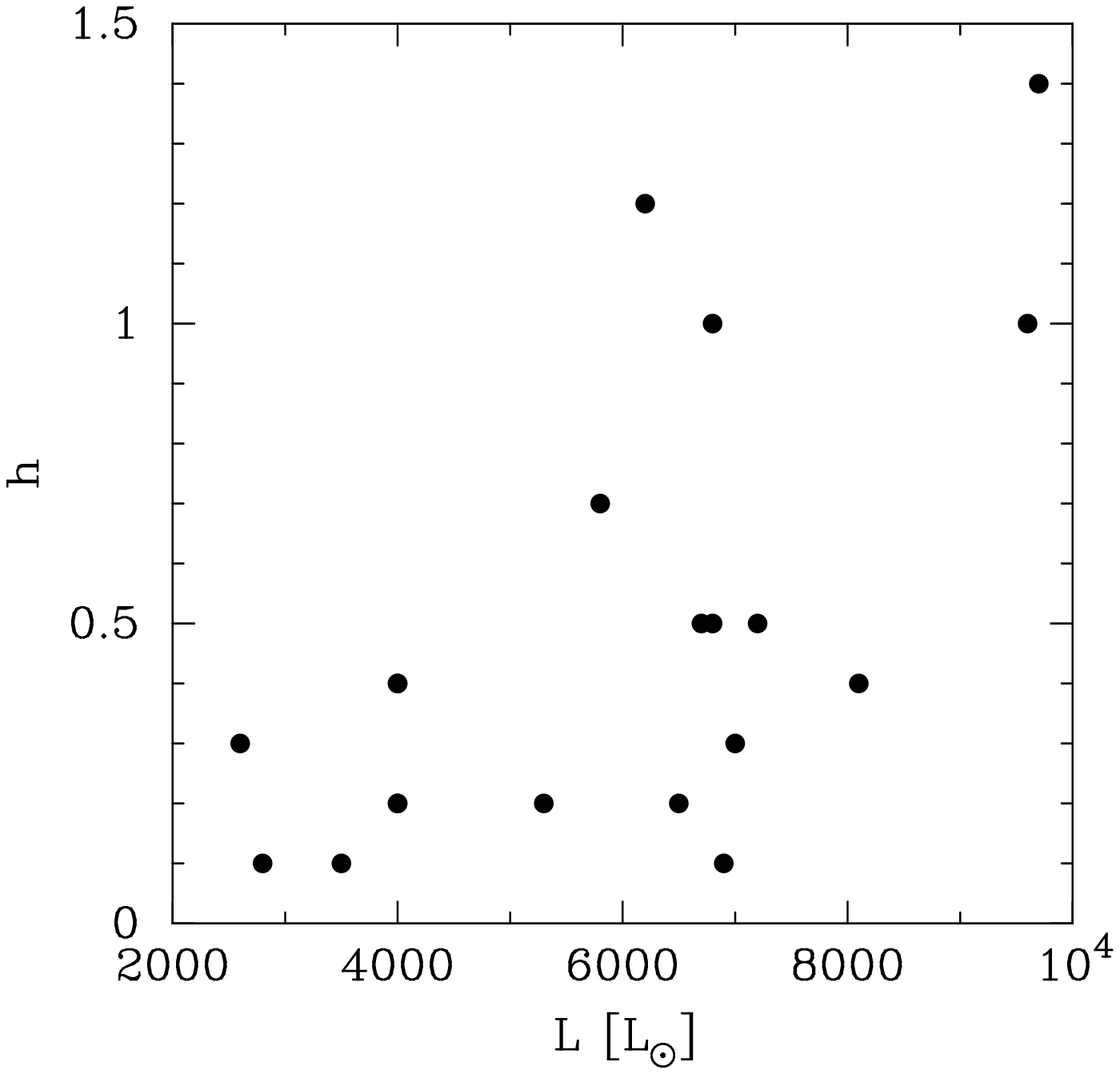}}
        \caption{The $h$-parameter derived from the radiative transfer analysis
        plotted against the adopted luminosity of the central star.
        Note that the data points at $L$$=$4000\,L$_{\sun}$ represent
        two stars each.}
        \label{h-para}
\end{figure}

\section{Model results}

\subsection{The fitting strategy}

With the assumptions made in the standard model there remains two principal
free parameters, the mass loss rate ($\dot{M}$) and the $h$-parameter.
These two parameters were allowed to vary simultaneously, $\dot{M}$
in steps of $\sim$10\% and $h$ in somewhat larger steps,
until the model with the smallest deviations from the observed
intensities was found.
The quality of a particular model with respect to the observational
constraints can be quantified using the chi-square statistic defined
as
\begin{equation}
\label{chi2_eq}
\chi^2 = \sum^{N}_{i=1} \frac{[I_{\mathrm{mod},i} -
I_{\mathrm{obs},i})]^2}{\sigma^2_{i}},
\end{equation}
where $I$ is the total integrated line intensity, $\sigma_i$ is the uncertainty
in observation $i$, and the summation is done over all independent
observations $N$. The errors in the observed intensities are 
generally dominated
by the calibation uncertainty of $\sim$20\%.
In some cases the line profiles were used to
discriminate between models. For each new $\dot{M}$ assigned, a consistent
envelope size ($r_{\mathrm p}$) was calculated based upon the results of
Mamon et~al. \cite*{Mamon88}.

It is found that the intensity ratios between the different lines are
sensitive to various parameters, and hence can be used to constrain them
(Sec.~\ref{mass-loss}).  Therefore, in order to accurately determine $h$, and
ultimately $\dot{M}$, multi-transition observations are needed.
Observed radial brightness profiles provide additional important
information.  For a limited number of the sources in our sample there
are enough
observational constraints that a reliable $h$-parameter may be
estimated.  These sources were first modelled, see
Table~\ref{coresult}.

\begin{figure*}
        \resizebox{\hsize}{!}{\includegraphics{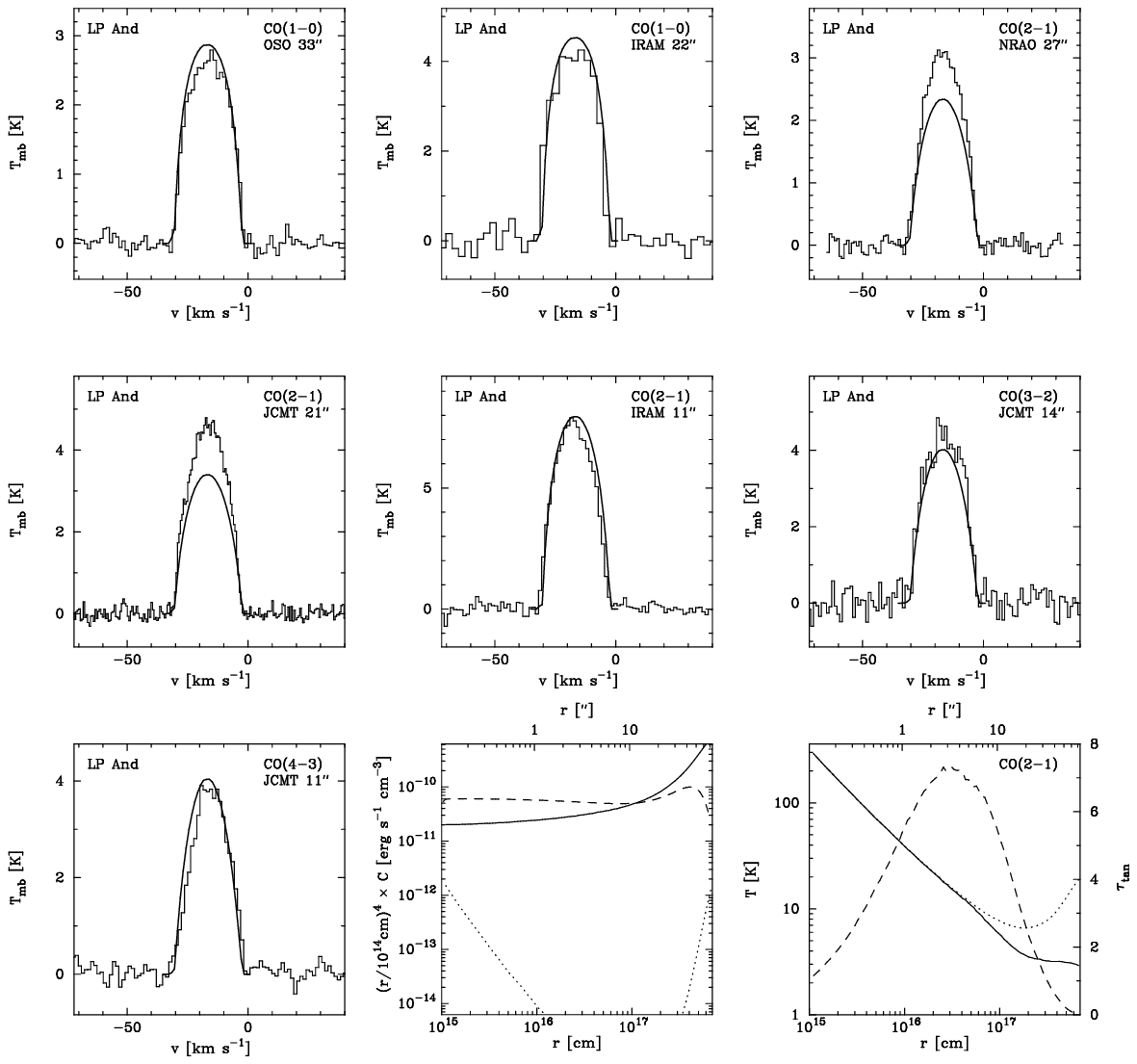}}
        \caption{Multi-transition CO  millimetre-wave line
        emission observed towards the high mass loss rate Mira variable
        \object{LP And}.  The observed spectra (histograms) have been
        overlayed with the model prediction (full line) using a mass loss rate
        of 1.5$\times$10$^{-5}$\,M$_\odot$\,yr$^{-1}$.  The transition,
        telescope used, and the corresponding beamsize, are indicated for each
        of the observations.
        {\em Cooling panel}: The full
        line represents adiabatic cooling; the dotted line gives the H$_2$
        cooling; and the dashed line is CO cooling (see text for details).
        {\em Temperature/optical-depth panel}: The dotted line shows
        the kinetic gas temperature derived from the energy balance equation.
        The full line gives the excitation temperature of the
        CO($J$=$2$$\rightarrow$$1$) transition.  The dashed lines give the
        tangential optical depth, $\tau_{\mbox{tan}}$, of the
        CO($J$=$2$$\rightarrow$$1$) transition.}

        \label{lpand}
\end{figure*}

\begin{figure*}
        \resizebox{\hsize}{!}{\includegraphics{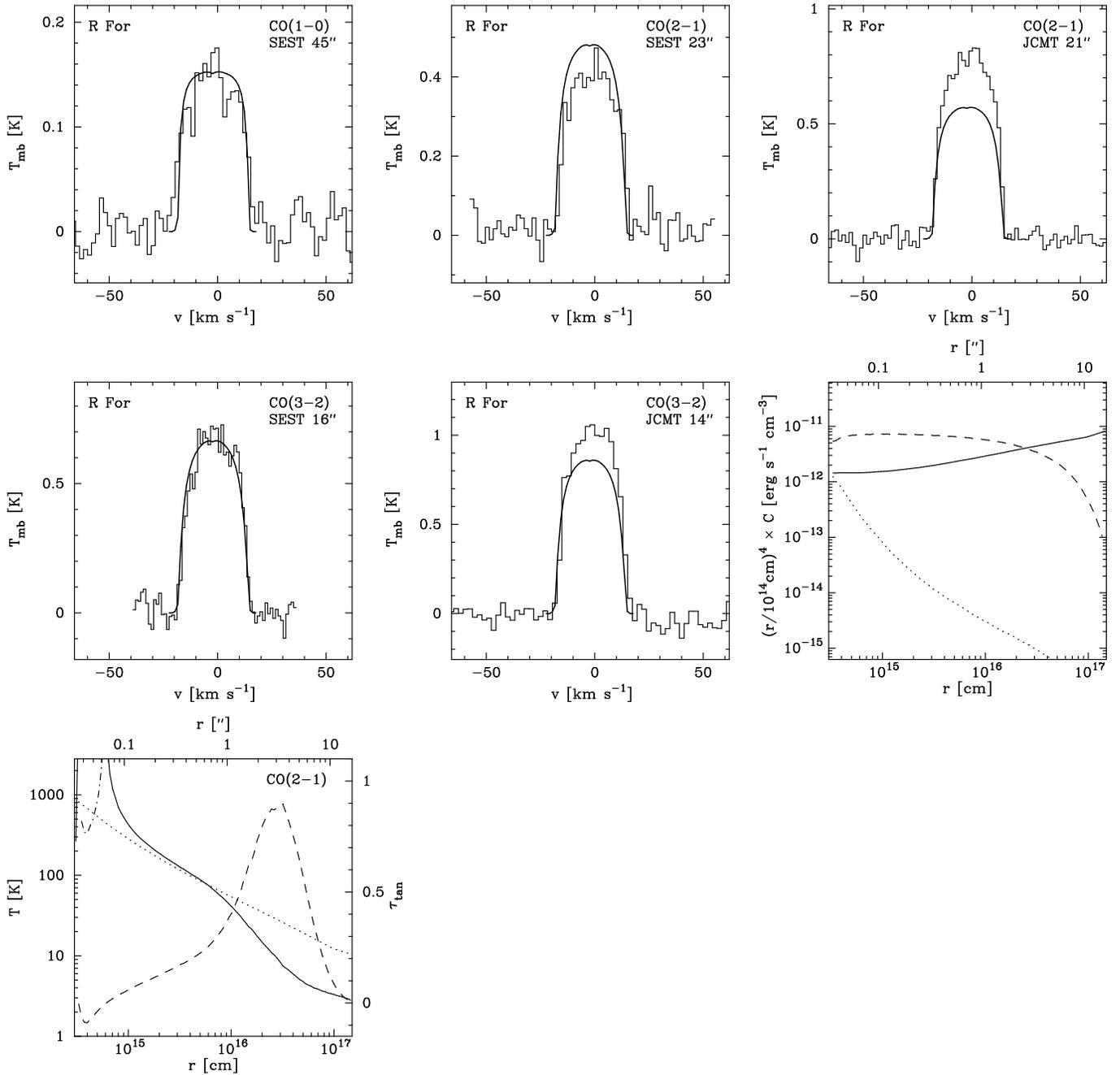}}
        \caption{Multi-transition CO  millimetre-wave line
        emission observed towards the bright carbon star \object{R~For}.
        The observed
        spectra (histograms) have been overlayed with the model prediction
        (full line) using a mass loss rate of
        1.3$\times$10$^{-6}$\,M$_\odot$\,yr$^{-1}$.  The transition, telescope
        used, and the corresponding beamsize, are indicated for each of the
        observations.
        {\em Cooling panel}: The full line represents adiabatic cooling; the
        dotted line gives the H$_2$ cooling; and the dashed line
        is CO cooling (see text for details).
        {\em Temperature/optical-depth panel}: The dotted line shows
        the kinetic gas temperature derived from the energy balance equation.
        The full line gives the excitation
        temperature of the CO($J$=$2$$\rightarrow$$1$) transition, and a
        dash-dot line indicates a negative excitation temperature, i.e., maser
        action. The dashed lines give the tangential optical depth,
        $\tau_{\mbox{tan}}$, of the CO($J$=$2$$\rightarrow$$1$) transition.}

        \label{rfor}
\end{figure*}

\begin{figure*}
        \resizebox{\hsize}{!}{\includegraphics{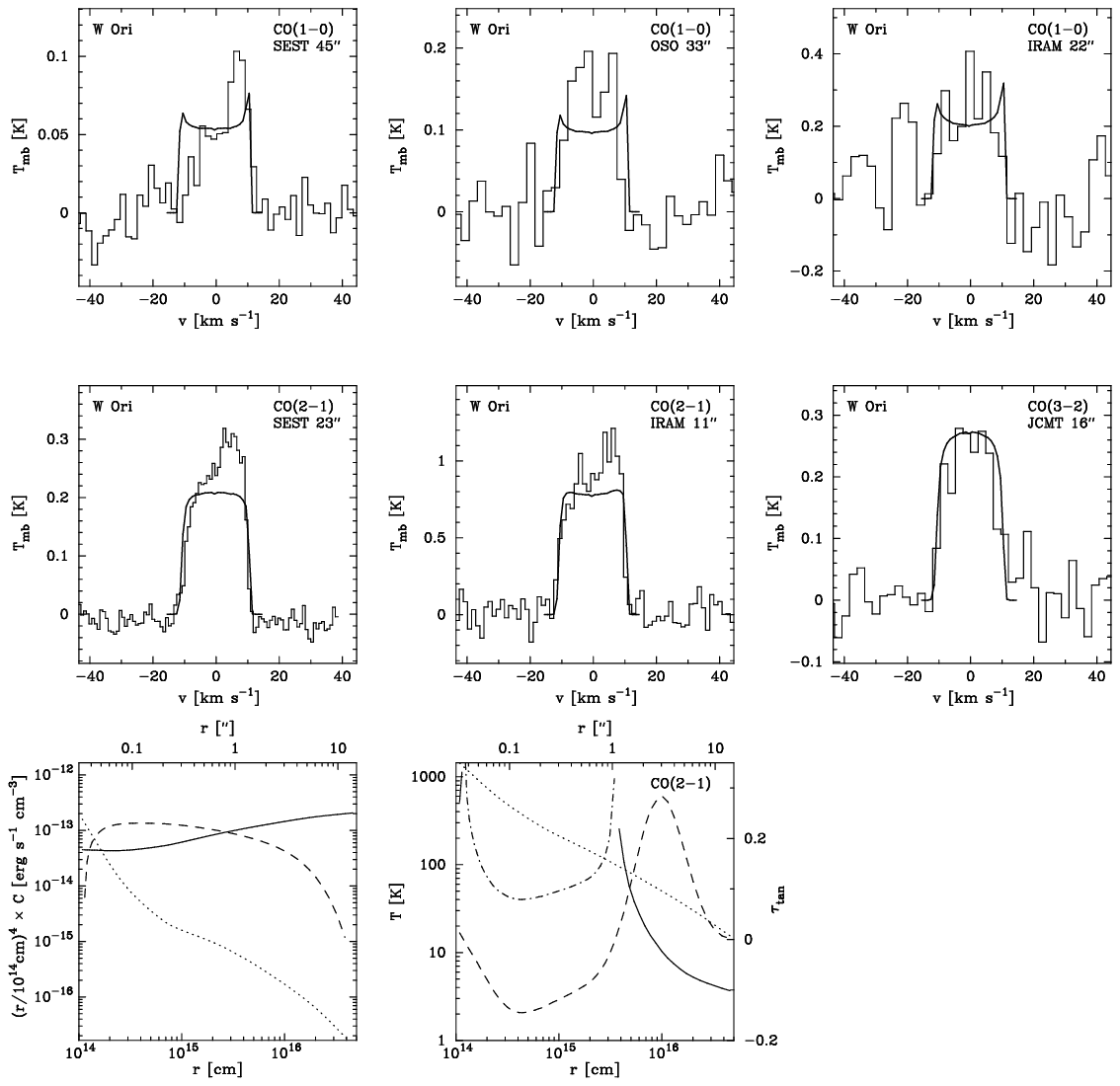}}
        \caption{Multi-transition CO  millimetre-wave line
        emission observed towards the bright carbon star \object{W Ori}.  The
        observed spectra (histograms) have been overlayed with the model
        prediction (full line) using a mass loss rate of
        7$\times$10$^{-8}$\,M$_\odot$\,yr$^{-1}$.  The transition, telescope
        used, and the corresponding beamsize, are indicated for each of the
        observations. The feature at $\sim$5\,km\,s$^{-1}$ may be of 
interstellar
        origin.
        The double-peaked profile in the  CO($J$=$1$$\rightarrow$$0$)
        model spectra is due to maser action and not due to spatial
        resolution effects.
        {\em Cooling panel}:
        The full
        line represents adiabatic cooling; the dotted line gives the H$_2$
        cooling; and the dashed line is CO cooling (see
        text for details).
        {\em Temperature/optical-depth panel}:
        The dotted line shows the kinetic gas temperature
        derived from the energy balance equation.  The full line gives the
        excitation temperature of the CO($J$=$2$$\rightarrow$$1$) transition,
        and a dash-dot line indicates a negative excitation temperature, i.e.,
        maser action.  The dashed lines give the tangential optical depth,
        $\tau_{\mbox{tan}}$, of the CO($J$=$2$$\rightarrow$$1$) transition.}

        \label{wori}
\end{figure*}

\subsection{The h-parameter}
\label{h-parameter}

The derived $h$-parameters, as a function of the
adopted luminosity of the central source, are shown in
Fig.~\ref{h-para}.  A division between low luminosity objects and
those with higher luminosities is evident.  For luminosities lower
than 6000\,L$_{\sun}$ the median value of the estimated
$h$-parameter is 0.2 (based on 9 sources), while for the higher
luminosities it is 0.5 (12 sources) (the J-type stars, i.e., stars
with a high content of $^{13}$CO, were not included in these estimates
since $^{13}$CO line cooling is not treated in the standard model).
The large scatter for the high luminosity sources is unfortunate,
since these are generally the sources where the derived mass loss
rates are particularly sensitive to the adopted value of $h$.  For the
majority of objects, where a reliable estimate of the $h$-parameter is not
possible, we have adopted a value of $h$$=$0.2 for the low luminosity
objects (below 6000\,L$_{\sun}$) and $h$$=$0.5 for the high luminosity
objects (above 6000\,L$_{\sun}$).

The $h$-parameter is related to the dust-to-gas mass ratio, $\Psi$ (see
Eq.~\ref{h}).  If we assume that the properties of a dust grain are
equal to those used to fit the line emission from \object{CW~Leo}, and
that they are the same for all stars, an $h$-parameter of 0.2 means
that $\Psi$$=$2$\times$10$^{-3}$.  Groenewegen et~al.
\cite*{Groenewegen98a} have modelled the dust emission towards a large
sample of carbon stars and find a value of $\Psi$ which is fairly
constant at 2.5$\times$10$^{-3}$ for stars with luminosities up to
about 7900\,L$_{\sun}$, and which increases drastically with
luminosity up to 0.01 and higher.  This is in good agreement with our
results and lends further support to our distinction between low- and
high-luminosity stars as regards the value of the $h$-parameter.
Furthermore, Hiriart \& Kwan \cite*{Hiriart00} derive, for 17 of our
sample stars, on the average $\Psi$$\sim$7$\times$10$^{-4}$.  The main
reason for this discrepancy is the difference in the
adopted dust parameters. In fact, it can be shown that for $h$$=$0.2
we get essentially the same heating term (cf. Eq.~\ref{H_dg})
in the energy balance equation as Hiriart \& Kwan \cite*{Hiriart00}.


\begin{figure*}
        \resizebox{\hsize}{!}{\includegraphics{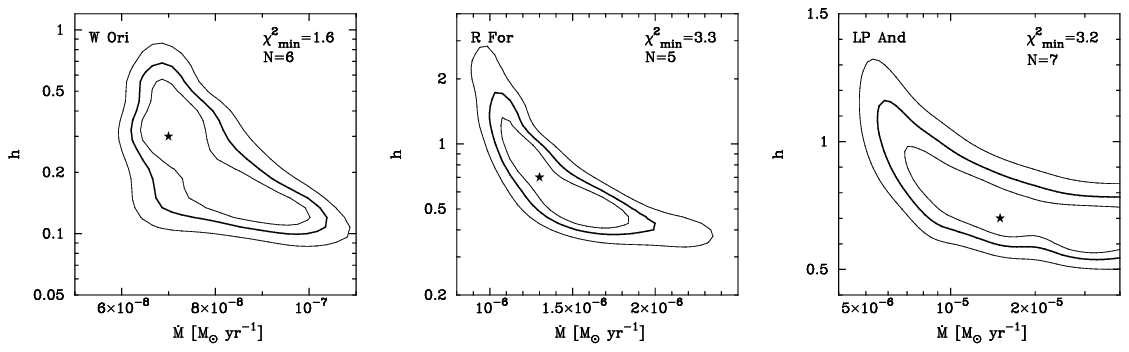}}
        \caption{$\chi^2$ maps showing the sensitivity of the model to the
        adjustable parameters, i.e., the mass loss rate ($\dot{M}$) and
        the $h$-parameter. Contour levels are drawn at
        $\chi^2_{\mathrm{min}}$$+$(1.0, 2.3, 4.6) with the middle contour
        (thick line) indicating the 68\% confidence level.
        The best-fit model, as reported in
        Tab.~\ref{coresult}, is indicated by a star and its corresponding
        $\chi^2$ and the number of observational constraints are also shown.}
        \label{chi2}
\end{figure*}

\begin{table*}
\caption[ ]{The effect on the integrated model intensity $I_{\mathrm{mod}}$
(in percent),
obtained by SEST and JCMT, 
due to changes in various parameters. In general, the accuracy of th observed
intensities lie at the $\pm$20\% level.}
\begin{flushleft}
\begin{tabular}{ccccccccccccccccc}
\hline
\noalign{\smallskip}
  & & &
 \multicolumn{4}{c}{\object{LP~And}} & &
 \multicolumn{4}{c}{\object{R~For}} & &
 \multicolumn{4}{c}{\object{W~Ori}} 
 \\ 
 \cline{4-7}
 \cline{9-12}
 \cline{14-17}
 \multicolumn{1}{c}{Parameter} &
 \multicolumn{1}{c}{Change} & &
 \multicolumn{1}{c}{1$-$0} & 
 \multicolumn{1}{c}{2$-$1} &
 \multicolumn{1}{c}{3$-$2} &
 \multicolumn{1}{c}{4$-$3} &
 &
 \multicolumn{1}{c}{1$-$0} & 
 \multicolumn{1}{c}{2$-$1} &
 \multicolumn{1}{c}{3$-$2} &
 \multicolumn{1}{c}{4$-$3} &
 &
 \multicolumn{1}{c}{1$-$0} & 
 \multicolumn{1}{c}{2$-$1} &
 \multicolumn{1}{c}{3$-$2} &
 \multicolumn{1}{c}{4$-$3}  
 \\
\noalign{\smallskip}
\hline
\noalign{\smallskip}
$\dot{M}^{1,2}$ & $+$50\%   &&           $+$25 &           $+$15 &           $+$15 &           $+$10 && $+$50 & $+$55 & $+$55 & $+$55 &&	   $-$10 & $+$15& $+$40 & $+$55\\
                & $-$33\%   &&           $-$25 &           $-$15 &           $-$15 &           $-$15 && $-$35 & $-$35 & $-$40 & $-$40 &&	   $-$40 & $-$30& $-$25 & $-$35\\
$\dot{M}$       & $+$50\%   &&           $+$15 & \phantom{0$+$}0 & \phantom{0$+$}0 & \phantom{0}$+$5 && $+$55 & $+$40 & $+$35 & $+$30 &&	   $+$25 & $+$30& $+$40 & $+$45\\
                & $-$33\%   &&           $-$15 & \phantom{0}$-$5 & \phantom{0}$-$5 & \phantom{0}$-$5 && $-$35 & $-$30 & $-$30 & $-$25 &&	   $-$40 & $-$25& $-$20 & $-$30\\
$L^1$           & $+$50\%   && \phantom{$+$0}0 & \phantom{$+$0}0 & \phantom{$+$0}0 & \phantom{$+$0}0 && \phantom{$+$0}0 & \phantom{0$-$}0& \phantom{0$-$}0& \phantom{$+$0}0&&  	 
$+$15 & $+$10 & \phantom{0}$+$5 & \phantom{0}$+$5\\
                & $-$33\%   && \phantom{$+$0}0 & \phantom{$+$0}0 & \phantom{$+$0}0 & \phantom{$+$0}0 && \phantom{0$-$}0 & \phantom{0}$-$5&
		\phantom{0$-$}0& \phantom{0$-$}0&&	   $-$20 & $-$10 & \phantom{0}$-$5 & \phantom{0}$-$5\\
$h$             & $+$50\%   &&  	 $+$25 &	   $+$35 &	     $+$35 &	       $+$40 && $+$10 & $+$15 & $+$20 & $+$25 && \phantom{$+$0}0 & \phantom{0}$+$5& $+$10
& $+$15\\
                & $-$33\%   &&  	 $-$20 &	   $-$25 &	     $-$30 &	       $-$30 && $-$10 & $-$15 & $-$20 & $-$20 && \phantom{0$-$}0 &           $-$10& $-$10
		& $-$15\\
$r_{\mathrm p}$ & $+$50\%   && \phantom{0}$+$5 & \phantom{$+$0}0 & \phantom{$+$0}0 & \phantom{$+$0}0 && $+$15& \phantom{0}$+$5 & \phantom{0}$+$5 & \phantom{0$+$}0 && 
          $+$20 & \phantom{0}$+$5 & \phantom{0}$+$5 & \phantom{0$+$}0\\
                & $-$33\%   && \phantom{0}$-$5 & \phantom{0$-$}0 & \phantom{$+$0}0 & \phantom{$+$0}0 && $-$20& \phantom{0}$-$5 & \phantom{$+$0}0 & \phantom{$+$0}0 &&	   
		$-$50 &    $-$30 & $-$10 & \phantom{0}$-$5\\
$r_{\mathrm i}$ & $\times$2 && \phantom{0}$+$5 & \phantom{$+$0}0 & \phantom{0$-$}0 & \phantom{0$-$}0 && \phantom{0}$+$5&\phantom{0}$+$5& \phantom{0}$+$5&\phantom{0$+$}0 && 
\phantom{0}$+$5 &      $+$15 & $+$15 & $+$10\\
$v_{\mathrm t}$ & $\times$2 &&           $+$10 & \phantom{0$+$}0 & \phantom{$+$0}0 & \phantom{$+$0}0 && $+$15 & $+$10 & \phantom{0}$+$5 & \phantom{0}$+$5 && \phantom{0$-$}0 &
 $+$25 & $+$40 & $+$30\\
\noalign{\smallskip}
\hline
\noalign{\smallskip}
  \noalign{$^1$ Same temperature structure as final model was used; $^2$Same envelope size as final model was used.}
\end{tabular}
\end{flushleft}
\label{error}
\end{table*}

\subsection{The mass loss rates}
\label{mass-loss}

We have managed to model reasonably well 61 of the 69 sample sources using our
standard model and fitting strategy defined above.
The reduced $\chi^2$ for these models is estimated from
\begin{equation}
\chi^2_{\mathrm{red}} = \frac{\chi^2_{\mathrm{min}}}{N-p},
\end{equation}
where $\chi^2_{\mathrm{min}}$ is obtained from Eq.~\ref{chi2_eq} for the
best fit model, and $p$ is the number of adjustable parameters (one or two)
in the model. Typically, $\chi^2_{\mathrm{red}}$$\sim$1$-$3 assuming a
calibration uncertainty of 20\%. The derived mass loss rates and
$h$-parameters are presented in Table~\ref{coresult}, where also the
velocities by which the CSEs expand, $v_{\mathrm e}$, and their
spatial extents, $r_{\mathrm p}$, are given.

The derived mass loss rates presented in Table~\ref{coresult}
span more than three orders of magnitude,
6.5$\times$10$^{-9}$\,M$_{\sun}$\,yr$^{-1}$
to 1.5$\times$10$^{-5}$\,M$_{\sun}$\,yr$^{-1}$,
over which the physical conditions of the CSEs vary considerably, posing a
challenge to the model.  The intensities, overall line shapes, and when
available radial brightness distributions, of the circumstellar CO
lines produced by the radiative transfer model generally agree well
with those observed.  This is illustrated here for three of our sample
stars \object{LP~And} (Fig.~\ref{lpand}), \object{R~For}
(Fig.~\ref{rfor}), and \object{W~Ori} (Fig.~\ref{wori}).  These
carbon stars span a large range in mass loss rate and serve to
illustrate the various physical conditions in these CSEs.

A full, detailed, error analysis of the estimated mass loss rates is not
possible due to the relatively large number of free parameters entering the
model.
Instead, we will vary some of the more important parameters in order to
illustrate the sensitivity of the model, and to be able to get a rough estimate
of the errors involved in the mass loss rate estimates.  The results of these
sensitivity tests are shown in Table~\ref{error} for our three example
stars.  
In addition, in Fig.~\ref{chi2} we present chi-square contour plots
for these stars, produced by varying the mass loss rate and the
$h$-parameter. This will give an estimate of the accuracy in the determination
of these two adjustable parameters when all other parameters are held fixed
(the size and shape of the CO envelope is however allowed to change in
accordance with Eqs~\ref{rout} and \ref{alpha}).  In Fig.~\ref{chi2} we
indicate various confidence levels with the contour at
$\chi^2_{\mathrm{min}}$$+$2.3 marking the 68\% confidence limit (i.e.,
the "1$\sigma$" level for two adjustable parameters).  The information
contained in the shape of the line profiles (to which the $\chi^2$
defined above is not very sensitive) have occasionally been used to further
constrain the most probable parameter space.

\object{LP~And} is a high mass loss rate Mira
variable where the excitation of $^{12}$CO is dominated by collisions.
Consequently, the line intensities are very sensitive to the temperature
structure, i.e., the $h$-parameter. It is interesting to note that
the assumed envelope size $r_{\mathrm p}$ does not significanly affect
the derived mass loss rate in the high mass loss rate regime.  The
reason for this being that the density is too low to excite the CO
molecules effectively (the lines are sub-thermally excited) in the
cool outer parts of the envelope (cf.  Fig.~\ref{lpand}), i.e., the
emission is excitation limited.  In the high mass loss rate regime the
line intensities are also insensitive to the adopted mass loss rate since an
increase in $\dot{M}$ leads to more cooling ($C$$\propto$$\dot{M}$
while $H$$\propto$$\dot{M}^{0.5}$), which compensates for the increase
of molecular density.  This "saturation"-effect of the line
intensities for high mass loss rates has been noted before in other
models where the cooling by CO is treated in a self-consistent manner
(Sahai 1990\nocite{Sahai90}; Kastner 1992\nocite{Kastner92}).
Additional sensitivity tests in the case of a high mass loss rate 
object (\object{CW~Leo}) can be found in Groenewegen (1994).

\object{W~Ori}, on the other hand, is a low mass loss rate object with low
optical depths in the lowest rotational transitions. In fact, some of
the low lying transitions are inverted over parts of the CSE,
i.e., maser action is taking place. In particular,
the $J$$=$$1$$\rightarrow$$0$ is inverted over the entire CSE producing a line
profile that is clearly double peaked, i.e., this is due to maser action
and not due to resolution effects.  The observations of
this transition are complicated by possible contamination by
interstellar lines, and the signal-to-noise ratio is not particularly good,
erasing any possible signs of maser action.  As expected the radiation
emitted by the central star plays an important role in the excitation.
The transitions exhibiting population inversion are particularly
sensitive (both the intensity and line profile) to the physical
conditions prevailing in the CSE. In this case where radiative
excitation is important the model becomes sensitive to the choice of
inner radius, $r_{\mathrm i}$, as well as tubulent velocity,
$v_{\mathrm t}$.  Here, also the size of the envelope is important
when determining the mass loss rate, i.e., the emission is
photodissociation limited (at least for the lowest transitions).
Generally, the line intensities scale roughly linearly with $\dot{M}$.

\object{R~For} presents an intermediate case between \object{LP~And} and
\object{W~Ori}. It is also a typical sample star in the sense that the
majority of the stars have envelope parameters that resemble those of
\object{R~For}. From Table~\ref{error} and Fig.~\ref{chi2}
we see that \object{R~For} shares properties of both the low and the high mass
loss rate star.  An intermediate mass loss rate model is generally
more sensitive to the temperature structure than the radiation field,
thus resembling the high mass loss rate objects.  However, as in the
low mass loss rate regime the line intensities scale roughly linearly
with $\dot{M}$.  We also note that the derived line intensities, especially
the $J$$=$$1$$\rightarrow$$0$ transition, start to become sensitive 
to the outer radius of the CO envelope. The exact
mass loss rate where the transition from photodissociation- to
excitation-limited emission occurs is hard to pinpoint, since it
depends on the envelope characteristics.  We have made tests which
show that for typical envelope parameters the transition lies at
$\sim$5$\times$10$^{-7}$\,M$_{\odot}$yr$^{-1}$ for the
$J$$=$$1$$\rightarrow$$0$ transition.  Higher $J$-lines sample hotter
and denser gas, located closer to the star, and are therefore less
sensitive to the choice of the envelope size.

In summary, we believe that within the adopted circumstellar model, the
estimated mass loss rates are accurate to about $\pm$50\% (neglecting
errors introduced by the uncertain CO abundance and the distance
estimates) when good observational contraints are available,
but one should keep in mind that the causes of the
uncertainty varies with the mass loss rate.

Another rough way to estimate the errors involved is to compare with
the results obtained from other self-consistent models.  When
comparing our mass loss rate estimate for \object{CW Leo} to those
obtained from detailed radiative transfer models (Kastner
1992\nocite{Kastner92}; Crosas \& Menten 1997\nocite{Crosas97};
Groenewegen et~al.\ 1998a\nocite{Groenewegen98b};
Skinner et~al.\ 1999\nocite{Skinner99}) we find a
very good agreement, within 20\,\%, when adjustments for differences
in $f_0$ and distance have been made.  Kastner \cite*{Kastner92} also
modelled the high mass loss rate object \object{RW LMi} (a.k.a.
\object{CIT 6}) and obtained a mass loss rate in excellent agreement
with our estimate (when corrected for the difference in distance).
Recently, Hiriart \& Kwan \cite*{Hiriart00} presented a method to
combine a circumstellar dust model with a CO model, keeping physical
quantities consistent between the models.  They present mass loss rate
estimates for 17 of our sample stars.  Based upon their Eq.~17 and
Table~6 we derive hydrogen mass loss rates for these stars that are,
on the average, 1.3 times larger than those derived from our radiative
transfer analysis (for an individual star, however, the discrepancy
can be as large as a factor of 3).

\begin{figure}
        \resizebox{\hsize}{!}{\includegraphics{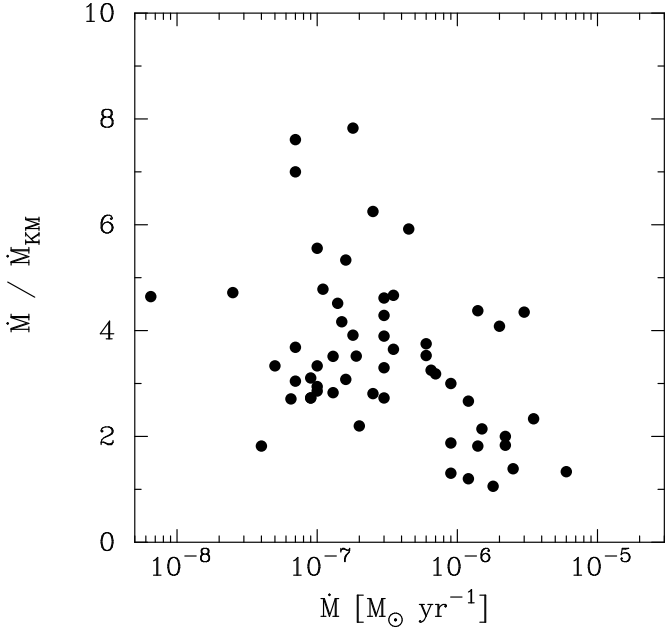}}
        \caption{The ratio of the estimated mass loss rates  from our
        radiative transfer analysis ($\dot{M}$) to those estimated from a
        analytical formula ($\dot{M}_{\mathrm{KM}}$) (see text for details).}
        \label{mdot-ratio}
\end{figure}

We also compare our derived mass loss rates with those obtained by
Olofsson et~al.\ (1993a)\nocite{Olofsson93a} using a formula based on
the work by Knapp \& Morris \cite*{Knapp85},
\begin{equation}
\dot{M}_{\mathrm{KM}} = 5.7\times 10^{-20} \frac{T_{\mathrm{mb}}
v^2_{\mathrm e} D^2
                \theta^2_{\mathrm{mb}}}{s(J) f_0^{0.85}}
	  \ \ \ {\mathrm M}_{\sun}\,{\mathrm{yr}}^{-1},
\label{mdot_simple}
\end{equation}
which relates the mass loss rate $\dot{M}_{\mathrm{KM}}$ to easily
determined observables ($T_{\mathrm{mb}}$ is given in Kelvins,
$v_{\mathrm e}$ in km\,s$^{-1}$, $D$ in pc, and $\theta_{\mathrm{mb}}$
in arcseconds). In Eq.~\ref{mdot_simple} $s(J)$ is a factor
that depends on the transition ($J$) used.  For
$J$$=$$1$$\rightarrow$$0$ $s(J)$$=$1 and for $J$$=$$2$$\rightarrow$$1$
$s(J)$$=$0.5 (see the discussion in Olofsson et~al.\
1993a\nocite{Olofsson93a}). When estimating the mass loss rates from
Eq.~\ref{mdot_simple} we have used the observed
$J$$=$$1$$\rightarrow$$0$ and $J$$=$$2$$\rightarrow$$1$ line emission,
and averaged the results.  This mass loss rate was compared to that
obtained from the radiative transfer analysis, Fig.~\ref{mdot-ratio}.
It is found that Eq.~\ref{mdot_simple} significantly underestimates
the mass loss rates when compared to those obtained from the radiative
transfer analysis by, on the average, a factor of about four.
However, considering the simplicity of Eq.~\ref{mdot_simple} the
agreement is remarkably good. There is also
a trend that the discrepancy increases with lower mass loss rate
(indeed Olofsson et~al.  suspected that their simple approach led to a
systematic underestimate of the mass loss rate that became worse the
lower the mass loss rate).
In this context it should be noted that Eq.~\ref{mdot_simple} was derived 
assuming a fixed CO envelope size of 3$\times$10$^{17}$\,cm,
which is appropriate for high mass loss rate objects.  In the high mass loss
rate regime the line intensities are not very sensitive to the exact
envelope size since the emission is excitation limited.  This is,
however, generally not the case for the optically bright carbon stars
in our sample, where the line intensities are sensitive to the size of
the CO envelope set by photodissociation.  A correction for the
envelope size [see Neri et al (1998)\nocite{Neri} and references
therein] would decrease the discrepancy seen in Fig.~\ref{mdot-ratio}
for the stars losing mass at lower rates. Taken at face values the
discrepancies between the mass loss rates derived here and those
derived by Olofsson et~al.\ (1993a)\nocite{Olofsson93a} are smaller,
but this is due to the fact that on the average Olofsson et~al.  used
larger distances.  This also means that the trend in the dust-to-gas
mass ratio, decreasing with the mass loss rate, that Olofsson et~al. 
reported was due to underestimated gas mass loss rates [see also
Hiriart \& Kwan (2000)].  Indeed, in Sec.~\ref{h-parameter} we find
the opposite behaviour, i.e., indications of a dust-to-gas mass ratio
that increases with the mass loss rate.

\begin{figure*}
        \resizebox{\hsize}{!}{\includegraphics{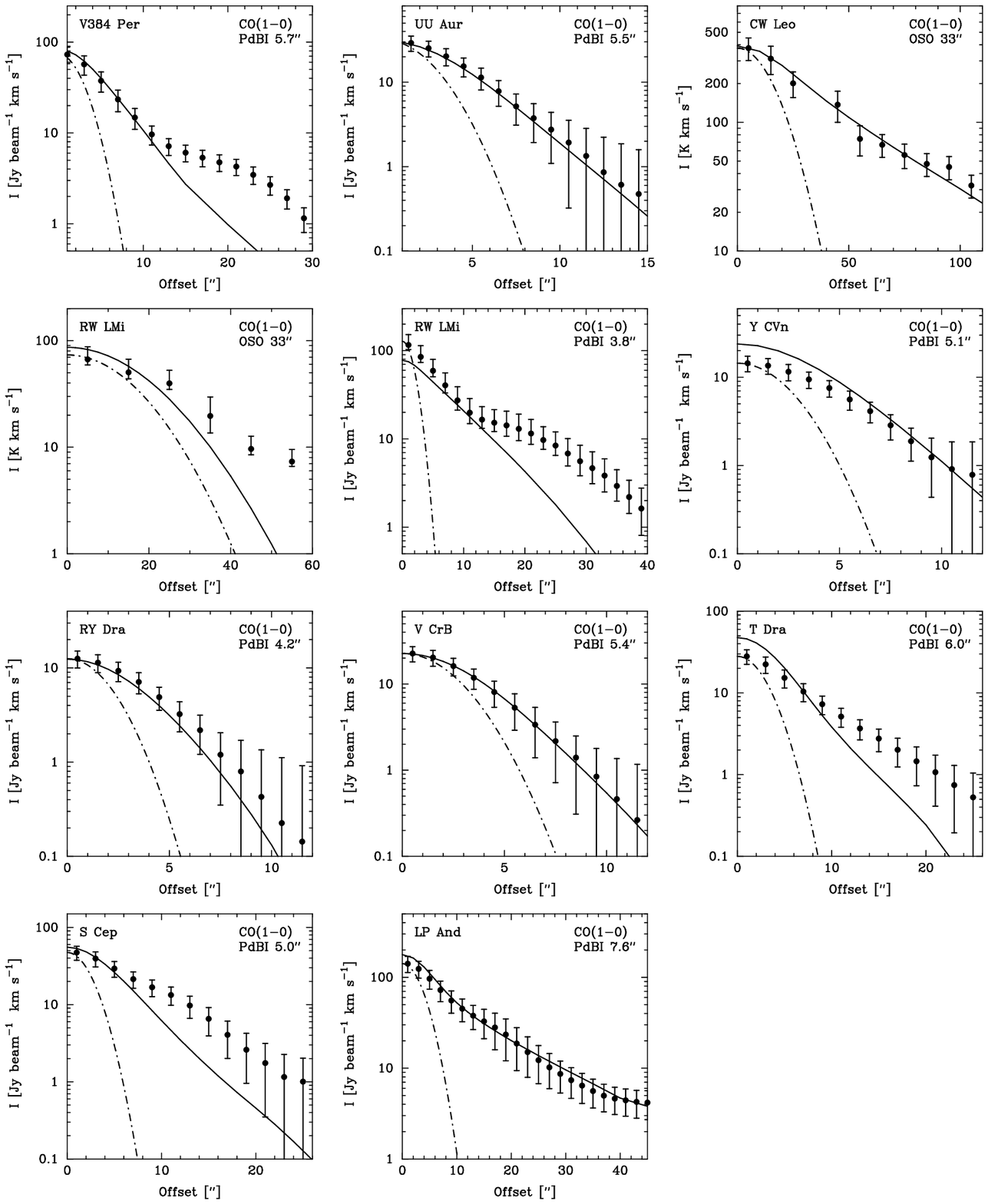}}
        \caption{Observed averaged radial brightness distributions overlayed
        by the results from the model, using the envelope sizes obtained from
        the chemical model by Mamon et al.  (1988), for the parameters
        presented in Tab.~\ref{coresult}.  The circular beam used in the
        radiative transfer calculation is indicated by the dot-dashed line.
        Observational errors include both statistical errors and a 20\%
        calibration uncertainty.}
        \label{radial}
\end{figure*}

\subsection{The CO envelope sizes}

The size of the CO envelope is an important parameter, and the derived mass
loss rate will depend on it to an extent that depends on the mass loss
rate (see discussion in Sec.~\ref{mass-loss}).
Neri et~al.\ \cite*{Neri} presented IRAM PdB interferometer (PDBI)
CO($J$$=$$1$$\rightarrow$$0$) maps of varying quality for 9 of our
sample stars. From their CLEANed brightness maps we have obtained the
averaged radial brightness distributions.
In addition, we have obtained two single-dish
CO($J$$=$$1$$\rightarrow$$0$) maps using the OSO 20\,m telescope.  In
Fig.~\ref{radial} we compare the observed radial brightness
distributions to those calculated from our model, which uses the Mamon et al.
\cite*{Mamon88} results to estimate the CO envelope size, for the
parameters presented in Tab.~\ref{coresult}.  We find that the estimated CO
envelope sizes are generally consistent with the observations.  In four cases
(\object{V384~Per}, \object{RW~LMi}, \object{T~Dra}, and
\object{S~Cep}) the calculated radial brightness distributions fail to
reproduce the observations.  At least in the case of \object{V384~Per}
and \object{RW~LMi} this appears to be due to deviations from a simple
$r^{-2}$ density law, in the sense that two distinct components are
clearly visible in the brightness maps.

We are not able to distinguish between the different CO envelope sizes
obtained by Mamon et al.
\cite*{Mamon88} and Doty \& Leung \cite*{Doty98} for the high mass
loss rate object \object{CW~Leo}.  This is simply because the emission
in this case is excitation limited with the external cool and tenuous
parts producing very little emission.

\subsection{The kinetic temperature structure}

As was demonstrated in Sec.~\ref{mass-loss} the derived mass loss rate can be
very sensitive to the kinetic temperature structure of the gas.
Adiabatic cooling alone results in $T(r)$$\propto$$r^{\delta}$, where
$\delta$$=$$-$4/3 (assuming $\gamma$$=$5/3), but
from Figs.~\ref{lpand}$-$\ref{wori} it is evident that other cooling processes
as well as heating mechanisms are important.
If CO line radiation is the dominant cooling mechanism and 
$n$$>$$n_{\mathrm{crit}}$ one would expect $T(r)$$\propto$$r^{-1}$ 
with a somewhat shallower decline if $n$$<$$n_{\mathrm{crit}}$ 
(see Doty \& Leung\ 1997\nocite{Doty97}).

Generally, it is not possible to assign a simple power law to the radial
behaviour of the temperature throughout the entire CSE.
However, in the case of \object{LP~And} we find $\delta$$\sim$$-$0.9 for
$r$$\lesssim$5$\times$10$^{16}$\,cm.  Further out in the cool, tenuous,
outer layers of this CSE, heating due to the photoelectric effect is
important and raises the temperature.  For \object{R~For} and
\object{W~Ori} we estimate, over the regions where the observed
emissions emanate, i.e., from
$\sim$1$\times$10$^{15}$ to $\sim$5$\times$10$^{16}$\,cm, $\delta$ to be
$\sim$$-$0.7 for both sources.  For comparison, Doty \& Leung
\cite*{Doty97} derive $\delta$$\sim$$-$1.0 for the high mass loss rate
object \object{CW~Leo}, while we obtain $\delta$$\sim$$-$0.9.

From the fact that $C$$\propto$$\dot{M}$ and
$H$$\propto$$\dot{M}^{0.5}$ one would expect high mass loss rate
objects to have significantly cooler envelopes than low mass loss rate
objects (e.g., Jura et al.\ 1988\nocite{Jura88}).
In our modelling we do not see
this trend.  The reason being that in addition $H$$\propto$$h$$(Lv_{\mathrm
e})^{3/2}$, where all of the parameters have a mass loss rate
dependence in the sense that their values are generally smaller for
the low mass loss rate objects.

\subsection{$^{13}$CO line cooling}
Carbon stars generally have $^{12}$C/$^{13}$C-ratios in the range 30$-$70
(Lambert et~al.\ 1986\nocite{Lambert86};
Sch\"{o}ier \& Olofsson 2000\nocite{Schoier00b}).
This means that for the large majority of all stars in our sample line
cooling by $^{13}$CO is insignificant, even when optical depth effects are
taken into account. However, there exists a small population of carbon stars,
the J-type stars, which have $^{12}$C/$^{13}$C-ratios of $\sim$3.  For
these stars $^{13}$CO line cooling could be important.

To test this we have included $^{13}$CO line cooling for one of our
sample stars, \object{Y~CVn}, a known J-type star. We have used the
$^{13}$CO model as presented in Sch\"{o}ier \& Olofsson \cite*{Schoier00b}
to calculate the the contribution from this isotopomer to the total amount
of cooling, and used this extra cooling term as input to the standard model.
We find that, after we have iterated between the $^{12}$CO and $^{13}$CO
models a few times, that $^{13}$CO line cooling is somewhat less effective
as a coolant than expected, based on the isotope ratio (probably due to
slightly different excitation conditions between the two molecules).
The  effect on the derived line intensities is small;
the largest effect is on the $J$$=$$3$$\rightarrow$$2$ line emission
which is lowered by $\sim$10\%. The changes in line intensity may
be compensated for by increasing the $h$-parameter.

Moreover, Ryde et~al. \cite*{Ryde99} present a model of the high mass loss
rate object \object{IRAS+15194-5115} using the radiative transfer code outlined
here. The $^{12}$C/$^{13}$C-ratio for this object was estimated to be
5.5, i.e., it resembles the J-type stars in this respect. It was found that
$^{13}$CO line cooling in this case was approximately one third of that of
$^{12}$CO (due to optical depth effects the $^{13}$CO line cooling is
in a relative sense more important).  The decrease in kinetic temperature
due to this extra cooling was compensated for by the fact that the
$h$-parameter (heating) had to be somewhat increased to produce the
observed line intensities.  Consequently, the derived mass loss rate was not
affected by introducing the $^{13}$CO line cooling.

\section{Deviations from the standard model}
For some stars, where there are enough observational constraints, it is
possible to establish deviations from the standard model.

\begin{table*}
\caption[ ]{Model results for the sources with known detached shells
(see text for details)}
 \resizebox{\hsize}{!}{
\begin{tabular}{llccccccccccc}
\hline
\noalign{\smallskip}
 &
 &
 &
 &
 \multicolumn{3}{c}{aCSE} & &
 \multicolumn{5}{c}{dCSE} \\
 \cline{5-7} \cline{9-13} 
 \noalign{\smallskip}
 \multicolumn{1}{c}{Source} & 
 \multicolumn{1}{c}{Var. type} &
 \multicolumn{1}{c}{$P$} &
 \multicolumn{1}{c}{$D$} &
 \multicolumn{1}{c}{${\dot M}$} & 
 \multicolumn{1}{c}{$v_{\mathrm e}$}& 
 \multicolumn{1}{c}{$r_{\mathrm p}$} & &
 \multicolumn{1}{c}{${\dot M}_{\mathrm s}$} & 
 \multicolumn{1}{c}{$v_{\mathrm s}$}& 
 \multicolumn{1}{c}{$r_{\mathrm s}$}&
 \multicolumn{1}{c}{$\Delta r_{\mathrm s}$}&
 \multicolumn{1}{c}{$M_{\mathrm s}$}
 \\
  &
  &
 \multicolumn{1}{c}{[days]} &
 \multicolumn{1}{c}{[pc]} &
 \multicolumn{1}{c}{$[\mathrm{M}_\odot\,\rm {yr}^{-1}]$} &
 \multicolumn{1}{c}{[km\,s$^{-1}$]} &
 \multicolumn{1}{c}{[cm]} & &
  \multicolumn{1}{c}{$[\mathrm{M}_\odot\,\rm {yr}^{-1}]$} &
 \multicolumn{1}{c}{[km\,s$^{-1}$]} &
 \multicolumn{1}{c}{[cm]} &
 \multicolumn{1}{c}{[cm]} &
 \multicolumn{1}{c}{$[\mathrm{M}_\odot]$}
 \\
\noalign{\smallskip}
\hline
\noalign{\smallskip}
\object{R Scl}$^1$  & SRb & 370 & 360 & & & && 4.0$\times$10$^{-6}$ & 16.5 & 7.0$\times$10$^{16}$& 7.0$\times$10$^{16}$ &
5.4$\times$10$^{-3}$ \\
\object{U Cam}$^2$  & SRb &     & 500 & 2.5$\times$10$^{-7}$ & 12.0 & 1.4$\times$10$^{16}$ && 9.0$\times$10$^{-6}$ & 23.0 &
5.5$\times$10$^{16}$& 1.1$\times$10$^{16}$  & 1.4$\times$10$^{-3}$ \\
\object{U Ant}$^1$  & Lb  &     & 260 & 3.0$\times$10$^{-8}$ & \phantom{0}4.5 & 1.4$\times$10$^{16}$ && 1.5$\times$10$^{-5}$ &19.5 &
1.6$\times$10$^{17}$& 2.0$\times$10$^{16}$  & 4.9$\times$10$^{-3}$ \\
\object{S Sct}$^1$  & SRb & 148 & 400 & 3.0$\times$10$^{-8}$ & \phantom{0}4.5 & 1.4$\times$10$^{16}$ && 4.0$\times$10$^{-5}$ & 17.0 &
4.0$\times$10$^{17}$& 2.0$\times$10$^{16}$ & 1.5$\times$10$^{-2}$  \\
\object{TT Cyg}$^3$ & SRb & 118 & 510 & 3.0$\times$10$^{-8}$ & \phantom{0}4.0 & 1.5$\times$10$^{16}$ && 1.5$\times$10$^{-5}$ & 12.5 &
2.7$\times$10$^{17}$& 1.9$\times$10$^{16}$ & 7.2$\times$10$^{-3}$ \\
\noalign{\smallskip}
\hline
\noalign{\smallskip}
\noalign{Model is presented in: $^1$ This paper; $^2$ Lindqvist et al. \cite*{Lindqvist99}; 
$^3$ Olofsson et al. \cite*{Olofsson00}.}
\end{tabular}}
\label{dCSE}
\end{table*}

\begin{table*}
\caption[ ]{CO modelling results, for three sources with known detached shells, compared to observations.}
\label{dCSE-model}
\begin{flushleft}
\begin{tabular}{lllcccccc}
\hline
 &
 &
 &
 \multicolumn{2}{c}{aCSE} & &
 \multicolumn{2}{c}{dCSE} & 
 \\
 \cline{4-5} \cline{7-8}
 \noalign{\smallskip}
 \multicolumn{1}{c}{Source} & 
 \multicolumn{1}{c}{Tel.} & 
 \multicolumn{1}{c}{Trans.}& 
 \multicolumn{1}{c}{$I_{\mathrm{obs}}$} &
 \multicolumn{1}{c}{$I_{\mathrm{mod}}$} & 
 &
 \multicolumn{1}{c}{$I_{\mathrm{obs}}$} &
 \multicolumn{1}{c}{$I_{\mathrm{mod}}$} &
 \multicolumn{1}{c}{Ref.} \\
  &
  &
  &
 \multicolumn{1}{c}{[K\,km\,s$^{-1}$]} &
 \multicolumn{1}{c}{[K\,km\,s$^{-1}$]} & 
  &
 \multicolumn{1}{c}{[K\,km\,s$^{-1}$]} &
 \multicolumn{1}{c}{[K\,km\,s$^{-1}$]} & \\
\hline
%
\object{R~Scl}$^{\mathrm a}$    & SEST & 1$-$0 &  	           &     & & \phantom{$<$}26.1 & 29.6 & 2\\
                  & IRAM & 1$-$0 &  	           &     & & \phantom{$<$}76.2 & 79.2 & 1\\
                  & SEST & 2$-$1 &  	           &     & & \phantom{$<$}47.8 & 49.8 & 1\\
		  & JCMT & 2$-$1 &  	           &     & & \phantom{$<$}50.6 & 54.8 & 3\\
		  & IRAM & 2$-$1 &  	           &     & & \phantom{$<$}67.7 & 72.0 & 1\\
		  & SEST & 3$-$2 &  	           &     & & \phantom{$<$}63.7 & 28.0 & 2\\
		  & JCMT & 3$-$2 &                 &     & & \phantom{$<$}62.9 & 30.5 & 3\\
\object{U~Ant}    & SEST & 1$-$0 &                 & 0.6 & & \phantom{$<$}10.0 &13.4 & 2\\
                  & SEST & 2$-$1 & 3.2\phantom{:}  & 2.8 & & \phantom{0$<$}7.6 & \phantom{0}7.3 & 2\\
		  & JCMT & 2$-$1 & 3.6\phantom{:}  & 3.3 & & \phantom{0$<$}7.7 & \phantom{0}7.2 & 3\\
		  & SEST & 3$-$2 & 3.6\phantom{:}  & 4.0 & & \phantom{0$<$}5.0 & \phantom{0}3.0 & 2\\
\object{S~Sct}    & SEST & 1$-$0 & 	           & 0.3 & & \phantom{0$<$}5.3 & \phantom{0}5.6 & 1\\
                  & IRAM & 1$-$0 & 2.5:            & 1.3 & & \phantom{0$<$}5.3 & \phantom{0}5.4 & 1\\
                  & SEST & 2$-$1 & 1.1\phantom{:}  & 1.3 & & \phantom{0$<$}2.7 & \phantom{0}2.9 & 2\\
		  & JCMT & 2$-$1 & 0.9\phantom{:}  & 1.3 & & \phantom{0$<$}2.4 & \phantom{0}3.0 & 3\\
		  & IRAM & 2$-$1 & 1.8:            & 5.4 & & \phantom{0$<$}2.8 & \phantom{0}2.8 & 1\\
		  & SEST & 3$-$2 & 2.1\phantom{:}  & 1.7 & & \phantom{0$<$}1.0 & \phantom{0}0.9 & 2\\
		  & JCMT & 3$-$2 & 2.4\phantom{:}  & 2.2 & & \phantom{0$<$}1.0 & \phantom{0}0.9 & 3\\	     
\hline
\noalign{\smallskip}
\noalign{$^{\mathrm a}$ For this source it is not possible to separate the 
contributions from an aCSE and a dCSE; 1.\ Olofsson et al. \cite*{Olofsson93a}; 
2. Olofsson et al. \cite*{OBEG}; 3. JCMT public archive}
\end{tabular}
\end{flushleft}
\end{table*}

\subsection{Detached CSEs}
In our sample there exist a number of stars that exhibit a 60\,$\mu$m
excess, placing them in the regions VIa and VIb in the IRAS
colour-colour diagram (see Olofsson et~al.\
1993a\nocite{Olofsson93a}), suggesting that their CSEs are abnormal in
some way.  Previous CO observations of five of these 60\,$\mu$m excess
stars (see Table~\ref{dCSE}) have revealed that their CSEs cannot have
been produced by a single smooth expanding wind, but rather suggested
a drastic modulation of the mass loss on a time scale as short as a
thousand years, which produced detached CSEs (dCSEs; Olofsson et~al.\
1996b\nocite{OBEG}).  Presently, these stars are losing matter at a
significantly lower rate, forming what is refered to as an attached
circumstellar envelope (aCSE).

To be able to put constraints on the mass loss and its
variation with time, high spatial resolution observations are needed.
Such observations have been carried out using the IRAM PdB
interferometer, with an angular resolution of $\sim$1$''$, for two of
the stars in Table~\ref{dCSE}, \object{U~Cam} and \object{TT~Cyg}.  It
was found that the dCSE around \object{U~Cam} is fairly young while
that around \object{TT~Cyg} is significantly older.  Both dCSEs show a
remarkable overall spherical symmetry, and they are geometrically very
thin.  The CO line emission towards these two sources, supplemented by
a detailed modelling using the radiative transfer code outlined in
this paper, were presented in Lindqvist et~al.  \cite*{Lindqvist99}
(\object{U~Cam}) and Olofsson et~al.\ (1998\nocite{Olofsson98},
2000\nocite{Olofsson00}) (\object{TT~Cyg}).

In this paper we have attempted to model the remaining three stars with known
detached shells (\object{R~Scl}, \object{S~Sct}, and \object{U~Ant}),
based upon the results obtained from the high angular
resolution observations described above. The results for both the
aCSEs and dCSEs around these stars are presented in Table~\ref{dCSE}
(where also the results for \object{TT~Cyg} and \object{U~Cam} are
included). Due to the limited observational constraints available,
the dCSEs around \object{S~Sct} and \object{U~Ant} were
modelled using the same geometrical thickness and kinetic temperature
structure as obtained for \object{TT~Cyg}
(Olofsson et~al.\ 2000\nocite{Olofsson00}).
For \object{R~Scl} we used the same geometrical thickness as adopted by
Olofsson et~al.\ 1996b\nocite{OBEG} and assumed a kinetic temperature decrasing
linearly with radius from a value of 50\,K at the inner shell radius.
The derived intensities from the model generally
agree well with those observed (Table~\ref{dCSE-model}).
However, we note that in the case of \object{R~Scl} we only get about half
of the observed $J$$=$$3$$\rightarrow$$2$ line emission from the
model, but here the aCSE could produce a significant amount of the
observed emission.  We find that the estimated shell masses, $M_{\rm s}$, are
relatively similar, but there is a trend of increasing shell mass with
age (the \object{R~Scl} shell mass may be overestimated due to the
difficulty in separating the aCSE and dCSE emission), which may
indicate that gas is being swept up.

The present mass loss rate was modelled for \object{U~Ant} and
\object{S~Sct} using the standard model. The results are presented in
Table~\ref{dCSE} and Table~\ref{dCSE-model}.  For \object{R~Scl} the
emission from the aCSE and dCSE cannot be separated.  Again, the
predicted line intensities agree, with one exception, well with those
observed, considering the uncertainties of the observations.  For
\object{S~Sct} the model of the aCSE fails to reproduce the IRAM
CO($J$$=$$2$$\rightarrow$$1$) observations, in contrast to the dCSE
model where the fit to the observations is excellent.  An explanation
to this inconsistency could be that there are pointing problems in
these observations (the IRAM observations are more sensitive to the
pointing due to its smaller beam), which would mainly affect the
emission observed from the aCSE. We find that the present mass loss
rates of the stars with the oldest shells are low, and that the
present mass loss rate of \object{U~Cam} (which has a much younger dCSE)
is significantly higher.

The central stars are semi-regular (SR) or irregular variables (Lb),
and they have low present mass loss rates.  They appear to have
been subject to drastic modulations of their mass loss rates over
relatively short periods of time (a few thousand years).  We estimate
that the CO dCSE emission is observable for $\sim$10$^4$ years
(Bergman et~al.\ 1993\nocite{Bergman93}).
If this mass ejection is a repeatable
phenomenon, and all our sample stars go through this phase, we
estimate that the time scale between mass ejections is about 10$^5$ years.
Therefore, the mass ejections may possibly be linked with the
helium-shell flashes (the process that leads to nuclear-processed
matter being transported to the surface of the star) predicted to
occur regularly in AGB stars with time intervals of the order of
10$^5$ years for the relevant masses
(e.g., Bl\"{o}cker 1995\nocite{Bloecker95}). 
We note that {\em all} carbon stars that have been found to have clearly 
detached gas shells are members of our sample, and that no M-type star with a 
detached CO shell has been detected.

\subsection{\object{V~Hya} and \object{TX~Psc}}

The sucessful modelling of the vast majority of our sample stars,
assuming a spherically symmetric CSE, suggests that axi- or non-symmetric mass
loss is not a common phenomenon among these optically bright carbon
stars.  However, in the case of \object{V~Hya} and \object{TX~Psc},
stars with complex radio line profiles, the possibility of bipolar
outflows has been considered (Heske et~al.\ 1989\nocite{Heske}; Kahane
et~al.\ 1996\nocite{Kahane96}).  Due to the complexity of these
outflows, no radiative transfer analysis was attempted here.

Kahane et~al. \cite*{Kahane96} suggest that \object{V~Hya}, in addition to the
high velocity bipolar outflow, has a slowly expanding spherical
component.  Using a 2D radiative transfer code, based upon the Sobolev
approximation, they derive a total mass loss rate of
$\sim$1.5$\times$10$^{-6}$\,M$_{\odot}$yr$^{-1}$ for this object.  The
authors further suggest that \object{V~Hya} is a transition object,
between carbon stars and PNe, having just developed a highly
axi-symmetric mass loss.  The slowly expanding spherical component is
then interpreted as the fossil CSE created during the AGB stage.
Knapp et~al.  \cite*{Knapp99}, however, argue, based upon the spectral
type, period, colours, and lack of ionizing radiation, that
\object{V~Hya} is still on the AGB. Moreover, they estimate the total
mass loss rate of this object to be
$\sim$4$\times$10$^{-5}$\,M$_{\odot}$yr$^{-1}$.  This, in addition to
the small dynamical age of the envelope, suggests that \object{V~Hya}
has entered its "superwind" phase.

The observed line profiles of the CO line emission towards \object{TX~Psc}
are somewhat peculiar (Heske et~al.\ 1989\nocite{Heske}; Olofsson et~al.\
1993a\nocite{Olofsson93a}).  As a consequence, the expansion velocity
derived from the observed radio line profiles range from
7.5 to 12.2\,km\,s$^{-1}$ (see Table~\ref{obs} and Olofsson
et~al.\ 1993a\nocite{Olofsson93a}) depending on the transition and the
telescope used.  Heske et~al.  \cite*{Heske}, mapped the circumstellar
CO line emission around this object, and interpreted the observed
asymmetries as being produced either by a bipolar outflow or a highly clumped
wind.  No detailed modelling exist for the CSE around \object{TX~Psc},
and its evolutionary status remains uncertain.  It might be in a state
similar to that of \object{V~Hya}.

\subsection{\object{DR~Ser}}

\object{DR~Ser} has the highest 60\,$\mu$m excess of all the sample
stars, which places it in the IRAS two-colour diagram among objects
having detached envelopes (see Table~\ref{dCSE}).  Indeed, Kerschbaum
\cite*{Kerschbaum99} suggests that this might be a detached shell
source based upon his modelling of the SED of this object.  However, the
CO single dish observations reveal no signs of double-peaked
profiles (confusion with interstellar lines complicates the
interpretation), which are indicative of a detached CO shell.  Moreover,
\object{DR~Ser} has a relatively low $^{12}$C/$^{13}$C-ratio of 6
(Abia \& Isern 1997\nocite{Abia97}).  Thus, it resembles the J-type
stars.

Attempting to model this source with a standard
model of its CSE fails to explain both the observed $J$$=$$1$$\rightarrow$$0$
and $J$$=$$2$$\rightarrow$$1$ line intensities in that they come out too
weak, by more than a factor of two.  This is due to the extremely cold
envelope, resulting from the adopted $h$-parameter of 0.2 (due to the
luminosity being lower than 6000\,L$_{\odot}$), in combination with a
relatively
high mass loss rate $\sim$2$\times$10$^{-6}$\,M$_{\odot}$yr$^{-1}$
required to maximize the line intensities.  Further increasing the
mass loss rate will make the envelope even cooler due to the increase
in CO line cooling.  If one instead assigns an $h$-parameter of $\sim$1
it is possible to obtain a reasonable fit to the data using a mass
loss rate of $\sim$5$\times$10$^{-6}$\,M$_{\odot}$yr$^{-1}$.  This
estimate must be regarded as highly uncertain.

\begin{figure*}
        \resizebox{\hsize}{!}{\includegraphics{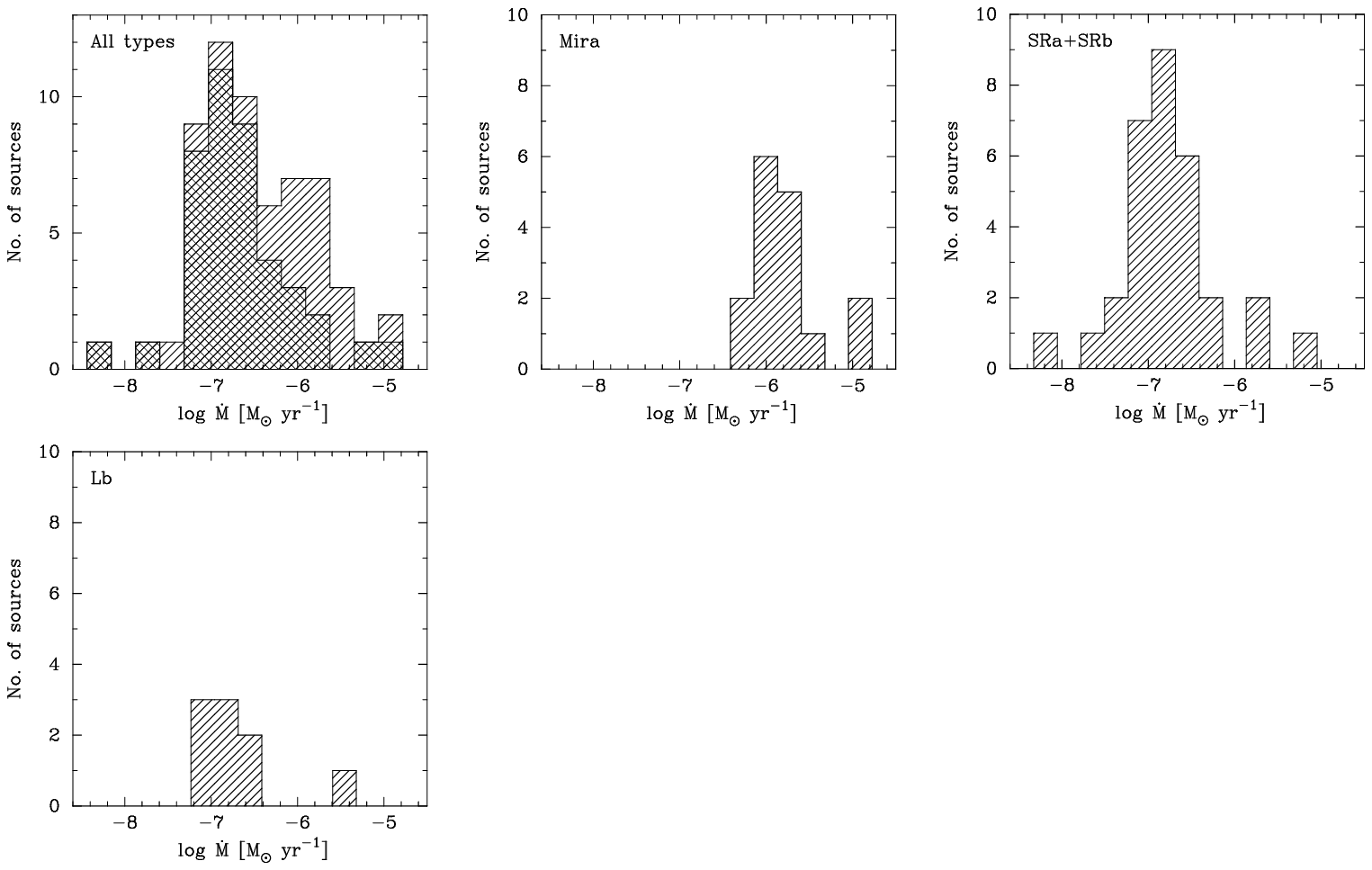}}
        \caption{Histograms showing the distribution of the mass loss rates
        for the whole sample (upper left panel), as well as subdivided into the
        different variability types
        (excluding \object{LP~And}).
        Stars with known detached shells have not been included in these plots.
        In the upper left panel the "cross-hatched"
        area indicates the mass loss rate distribution for the stars within
        500\,pc of the Sun, i.e., the complete sample.}
        \label{distr_all}
\end{figure*}

\section{Discussion}

\subsection{The mass loss rate distribution}

Fig.~\ref{distr_all} shows the distribution of the mass loss rates obtained
from the radiative transfer model for all sources (also shown
is the distribution for the stars within 500\,pc of the Sun), as well
as divided
into variability groups (note that objects with known detached shells
have not been included in this analysis).  Since the statistics of the
SRa stars are poor we have chosen to group them together with the SRb
stars.  The mass loss rate distribution for all stars is sharply peaked
around the median mass loss rate of 
2.8$\times$10$^{-7}$\,M$_{\sun}$\,yr$^{-1}$ (the median mass loss rate for the
stars within 500\,pc of the Sun is 1.6$\times$10$^{-7}$\,M$_{\sun}$\,yr$^{-1}$).
It is interesting to note
that this is roughly the rate at which the core mass is expected to
grow due to nuclear burning (Sch\"{o}nberner 
1983\nocite{Schoenberner83}).  Mira
variables generally have larger mass loss rates than other variability
types, while irregular variables and semiregulars appear to have very
similar mass loss rate characteristics.

Based on the mass loss rate distribution for the sample of stars
within 500\,pc, which we believe to be close to complete and for
which we have detected all sources, we are able to draw some general
conclusions.  The sharp decline at mass loss rates below
$\sim$5$\times$10$^{-8}$\,M$_{\odot}$yr$^{-1}$ is very likely real. 
Netzer \& Elitzur \cite*{Netzer93} estimate that a mass loss rate
in excess of $\sim$10$^{-7}$\,M$_{\sun}$\,yr$^{-1}$ is required
to get a dust-driven wind.  The exact limit is, however, sensitive to
the adopted stellar and dust parameters. The drastically decreasing
number of high mass loss rate objects is also real, although our
selection criterion bias against these objects, and can be explained
by the fact that carbon stars of the type discussed here, i.e., mainly
low mass ones (Claussen et~al.\ 1987\nocite{Claussen87}), only for a
limited time, or possibly never, reach high mass loss rates.

\begin{figure}
        \resizebox{\hsize}{!}{\includegraphics{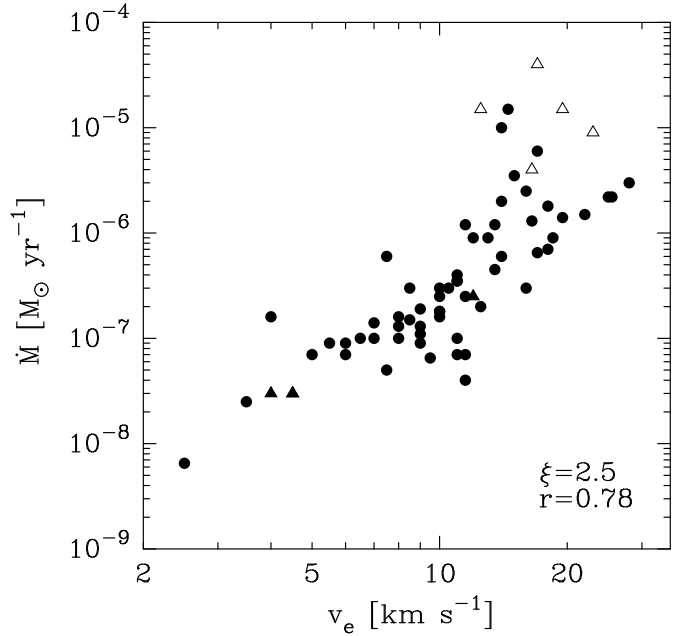}}
        \caption{The derived mass loss rate plotted against the gas expansion
        velocity of the CSE. Sources with known detached shells are indicated
        with triangles (open triangles represent dCSEs, while filled ones
        represent aCSEs).  $\dot{M}$ scales as $v_{\mathrm e}^{\xi}$, with
        $\xi$$=$2.5.  Also shown is the correlation coefficient $r$.}
        \label{mdot-ve}
\end{figure}

\subsection{Mass loss and envelope kinematics}

The $v_{\mathrm e}$-distribution for this sample of stars has already
been shown and discussed by Olofsson et al.  \cite*{Olofsson93a}.
However, a new comparison between the mass loss characteristics
$\dot{M}$ and $v_{\mathrm e}$ is warranted considering the more
reliable mass loss rates obtained in this paper, Fig.~\ref{mdot-ve}.
We find a clear trend that $v_{\mathrm e}$ increases with $\dot{M}$,
$v_{\mathrm e}$$\propto$$\dot{M}^{0.40}$ (with a correlation
coefficient of 0.78).  In comparison, Olofsson et~al.
\cite*{Olofsson93a} derived $v_{\mathrm e}$$\propto$$\dot{M}^{0.53}$,
but their mass loss rates were calibrated using Eq.~\ref{mdot_simple},
which includes a $v_{\mathrm e}^2$-dependence.  The correlation is
much tighter than that obtained by Olofsson el al.
\cite*{Olofsson93a}, which is reassuring.  Thus, the mass loss
mechanism operates such that mass loss rate and expansion velocity
increase together.  However, the scatter appears larger than the
uncertainties in the estimates, and so the mechanism also produces
widely different mass loss rates for a given expansion velocity.  At
high mass loss rates the scatter is larger with a possible divison
into objects with high mass loss rates but only moderately high
velocities, and objects with moderately high mass loss rates but high
velocities.  We note also that for the detached shell sources the mass
loss rates and expansion velocities for the two mass loss epochs
follow the general trend.

Habing et~al. \cite*{Habing94} studied the momentum transfer from the photons,
via the dust, to the gas in the CSEs of AGB stars. They found that for
mass loss rates close to the minimum mass loss rate for a dust-driven wind,
$\sim$3$\times$10$^{-8}$\,M$_{\sun}$\,yr$^{-1}$, $v_{\mathrm e}$
increases linearly with $\dot{M}$, while for mass loss rates above
$\sim$10$^{-6}$\,M$_{\sun}$\,yr$^{-1}$ the dependence weakens
considerably.  At 10$^{-5}$\,M$_{\sun}$\,yr$^{-1}$ they found
$v_{\mathrm e}$$\propto$$\dot{M}^{0.04}$. Habing et al. 
attribute the increase in $v_{\mathrm e}$ with mass loss rate (for low
mass loss rates) to a higher efficiency in the coupling between gas
and dust.  This can explain the observed behaviour in
Fig.~\ref{mdot-ve} for mass loss rates below about
10$^{-6}$\,M$_{\sun}$\,yr$^{-1}$, but for the higher mass loss rates it
appears that a dependence of $v_{\mathrm e}$ (and $\dot{M}$) on
luminosity gives the most reasonable explanation, see
Sect.~\ref{s:stellarprop}.

\subsection{Dependence on stellar properties}
\label{s:stellarprop}

\begin{figure*}
        \resizebox{\hsize}{!}{\includegraphics{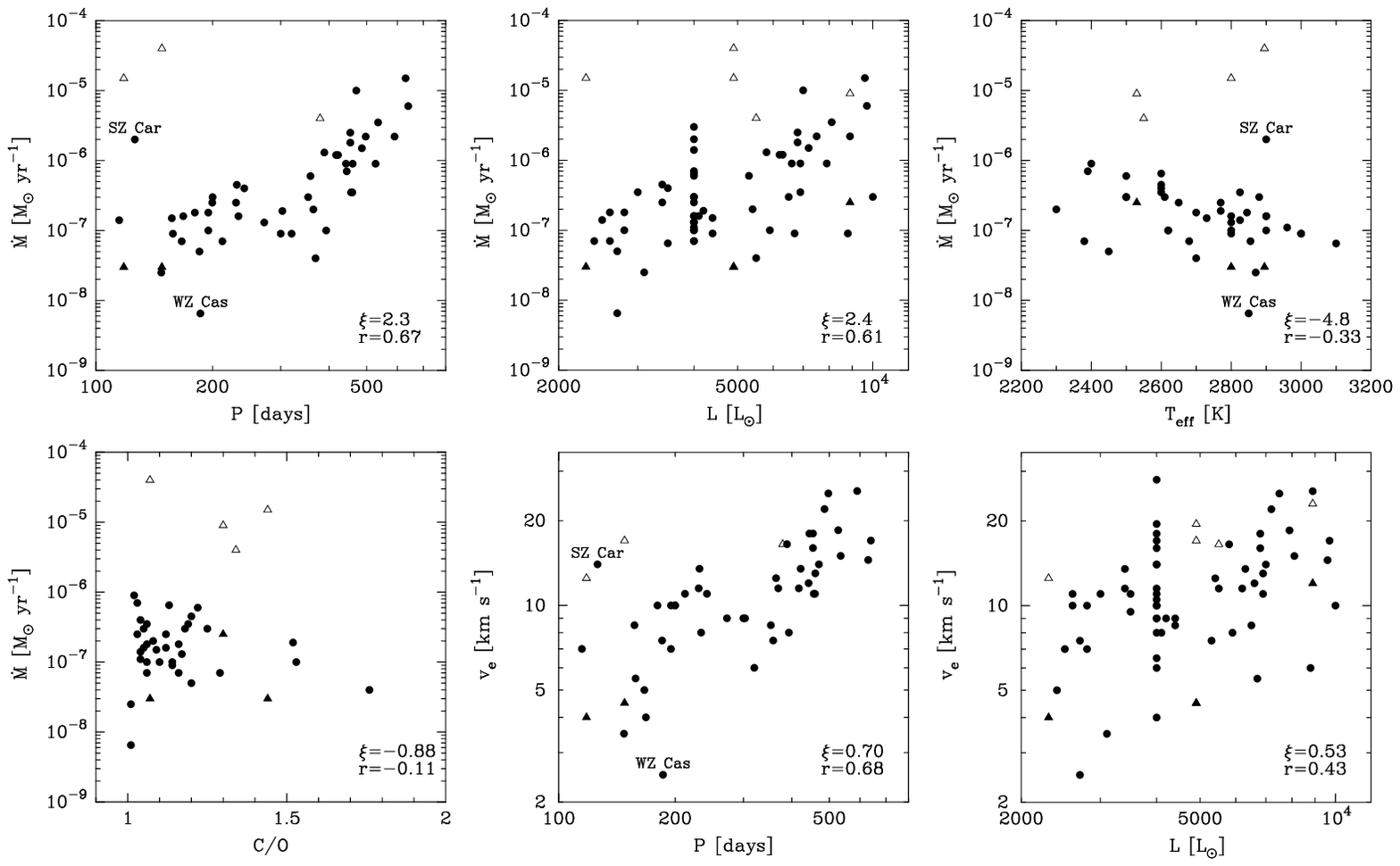}}
        \caption{The derived mass loss rate plotted against the luminosity
        ($L$), pulsational
        period ($P$), effective temperature ($T_{\mathrm{eff}}$), and
        the photospheric C/O-ratio of the star.
        The measured expansion velocity is plotted gainst the period and
        the luminosity.  Sources with known detached shells are indicated
        with triangles (open triangles represent dCSEs, while filled ones
        represent aCSEs).  Also shown are the power law index $\xi$, assuming
        the ordinate to scale with the abscissa, as well as the correlation
        coefficient $r$ (the detached shell sources are not included in the
        fit).}
        \label{mdot-stellar}
\end{figure*}

In Fig.~\ref{mdot-stellar} we plot the circumstellar characteristics,
$\dot{M}$ and $v_{\mathrm e}$, against the stellar characteristics luminosity
($L$), period ($P$), effective temperature ($T_{\mathrm{eff}}$), and the
photospheric C/O-ratio.  The effective temperatures and C/O-ratios used in
this analysis are those presented in Olofsson et~al.
\cite*{Olofsson93b}.  We have looked for dependences assuming
that the ordinate scales as the abscissa to the power of $\xi$, and a normal
correlation coefficient $r$ was calculated as an estimate of the quality of
the fit, see Fig.~\ref{mdot-stellar}.  The uncertainties in the 
estimated quantities are of the
order: $\pm$50\% ($\dot{M}$), $\pm$2\,km\,s$^{-1}$ ($v_{\mathrm e}$),
$\pm$10\% ($P$; but some periods may be poorly determined), a factor
of 2 ($L$), and $\pm$200\,K ($T_{\mathrm{eff}}$).

Clearly, both the mass loss rate and the expansion velocity
increase with the pulsation period of the star.  There is a weaker
trend with the luminosity of the star.  The latter dependence is not
completely independent of the former, since some of the luminosities
are estimated from a $P-L$ relation.  The existence of a $P-L$
relation is usually attributed to a distribution in mass (Jones et al. 
1994\nocite{Jones94}), i.e., the higher the mass the longer the period
and the higher the luminosity.  When looked at in detail, at least the
apparent mass loss rate dependence on period may be attributed to a
change from semiregular pulsation at short periods to regular
pulsations at longer periods.  We may therefore infer that for the
mass lass rate it is not clear whether it is the regularity of the
pulsation or the luminosity that causes the increase with period.  For
the expansion velocity the increase with period may be a combined
effect of increasing mass loss rate with period (for low mass loss
rates) and an increase in luminosity for the longer periods, e.g.,
Habing et~al.  \cite*{Habing94} derive $v_{\mathrm
e}$$\propto$$L^{0.35}$ in their dust-driven wind model.  In addition,
there is evidence of a weak trend in the sense that higher mass loss
rate objects have lower effective temperatures.  These correlations
are all consistent with a dust-driven wind, where the pulsation may
play an important role. There appears to be no correlation between
the mass loss rate and the C/O-ratio, which is surprising considering
that the dust-to-gas mass ratio in a C-rich CSE should be sensitively
dependent on this.

The present mass loss rates for the detached shell sources are typical
for their periods but somewhat low for their luminosities.  However,
during the formation of the dCSEs the mass loss rate must have been
atypically high for the present periods and luminosities. We also find
that the expansion velocities of the dCSEs lie at the very high end of
expansion velocities found for other stars with the same
luminosity, indicating that these stars had higher luminosities during
the shell ejection.

Two other stars, \object{SZ~Car} and \object{WZ~Cas}, stand out in
these plots.  The properties of \object{SZ~Car} resembles the stars
with known dCSEs, but the CO line profiles give no indication of a
detached shell.  \object{WZ~Cas} has by far the lowest estimated mass
loss rate in the sample, as well as the lowest C/O-ratio, 1.01, which
classifies it as an SC-star. It is also a Li-rich J-star (Abia \& Isern\
1997\nocite{Abia97}).

\subsection{Enrichment of the ISM}

Carbon stars on the AGB are important in returning processed gas to the
interstellar medium (ISM).  The total mass loss rate of carbon stars
in the Galaxy is obtained from
\begin{equation}
\label{totmassloss}
\dot{M}_{\mathrm{Gal}} = \int_0^{\infty} 2\pi R\,
\mbox{$\Sigma$$<$$\dot{M}$$>$}\, dR
\end{equation}
where $\Sigma$ is the surface density of carbon stars, $<$$\dot{M}$$>$
is the mean mass loss rate of the carbon stars.  According to
Guglielmo et~al. \cite*{Guglielmo98} the infrared carbon stars have a roughly
constant surface density out to the galactocentric distance of the
Sun, after which it follows an exponential decline with a scale length
of $\sim$2.5\,kpc.  We approximate this by assuming a constant value,
$\Sigma_0$ (the surface density in the solar neighborhood), to 10\,kpc
and zero beyond this galactocentric distance.

Based on the complete sample, i.e., stars within 500\,pc
from the Sun, we estimate $\Sigma_0$$<$$\dot{M}$$>$ to be
$\sim$1.7$\times$10$^{-10}$\,M$_{\sun}$\,yr$^{-1}$\,pc$^{-2}$, where
we have included also helium.  This estimate is very sensitive to the
number of high mass loss rate objects found within 500\,pc, e.g., the
high mass loss rate object \object{CW~Leo} contributes almost half of
our estimate of the total mass returned to the ISM. Our estimate of
the rate at which matter is returned to the ISM by carbon stars is
consistent with previous estimates (considering the large
uncertainties), e.g., Knapp \& Morris \cite*{Knapp85} and Jura \&
Kleinmann \cite*{Jura89} derive values of
$\sim$2$\times$10$^{-10}$\,M$_{\sun}$\,yr$^{-1}$\,pc$^{-2}$ and
$\sim$1.5$\times$10$^{-10}$\,M$_{\sun}$\,yr$^{-1}$\,pc$^{-2}$,
respectively.  Using Eq.~\ref{totmassloss} the annual return of
matter to the ISM in the Galaxy
is estimated to be $\sim$0.05\,M$_{\sun}$ for the carbon stars
considered here.  It is quite possible that a larger mass return from carbon
stars is obtained during their final evolution on the AGB, a so called
superwind phase, as is indicated by the estimated values for infrared
carbon stars, $\sim$0.5\,M$_{\sun}$\,yr$^{-1}$
(Epchtein et~al.\ 1990\nocite{Epchtein90}; Guglielmo et~al.\ 1993
\nocite{Guglielmo93}), and PNe,
$\sim$0.3\,M$_{\sun}$\,yr$^{-1}$ (Gustafsson et~al.\
1999\nocite{Gustafsson99}).  This confirms the importance of carbon
stars for the cosmic gas cycle in galaxies.  By comparison, high mass
stars are estimated to contribute about 0.04\,M$_{\sun}$\,yr$^{-1}$
(Gustafsson et~al.\ 1999\nocite{Gustafsson99}).

We estimate the contribution from carbon stars to the carbon enrichment
of the ISM to be $\sim$0.5$\times$10$^{-4}$\,M$_{\sun}$\,yr$^{-1}$
during the AGB stage (if the subsequent superwind phase is included this value
may increase to $\sim$5$\times$10$^{-4}$\,M$_{\sun}$\,yr$^{-1}$).  This
corroborates the conclusions by Gustafsson et~al.
\cite*{Gustafsson99} that `normal' carbon stars are not important for
the total carbon budget of the Galaxy.  According to these authors the
main contributors should instead be high mass stars in the Wolf-Rayet
stage, annually supplying the Galaxy with $\sim$0.01\,M$_{\sun}$ of
carbon.  This is roughly what is required to produce the
$\sim$10$^{-3}$$\times$10$^{11}$\,M$_{\sun}$ of carbon present in the
Milky Way over a period of 10$^{10}$ years.

\section{Conclusions}

We have developed a detailed radiative transfer code to model
circumstellar molecular line emission.  The code also solves for the energy
balance equation of the gas.  It is found that the mass loss rate
determination for low mass loss rate objects depends crucially on a
number of assumptions in the CSE model, except on the dust
properties, since the CO molecules are radiatively excited.  On the
other hand, the mass loss rate estimate for a high mass loss rate
object depends essentially only on the temperature structure, and
hence the uncertain dust parameters.  In addition, for such an object
the CO emission saturates, and becomes less useful as a mass loss rate
measure.  We also find that different lines respond differently to
changes in the various parameters.  Therefore, a reliable mass loss
rate determination requires, in addition to a detailed radiative
transfer analysis, good observational constraints in the form of
multi-transition observations and radial brightness distributions.

This model has been applied to CO radio line observations of a large
sample of optically bright N-type and J-type carbon stars on the AGB (69
objects).  The sample is reasonably complete out to about 1\,kpc, and
all stars (41) within $\sim$500\,pc of the Sun have been detected in
circumstellar CO line emission.  The derived mass loss rates span
almost four orders of magnitude,
$\sim$5$\times$10$^{-9}$\,M$_{\odot}$yr$^{-1}$ to
$\sim$2$\times$10$^{-5}$\,M$_{\odot}$yr$^{-1}$, over which the
physical conditions of the CSEs vary considerably.  The fact that the
model can be succesfully applied over such a wide range of
environments gives us confidence in the results, even though we are
aware of the fact that some assumptions are poorly constrained.  The
large majority of the stars have mass loss rates in a narrow range
centered at $\sim$3$\times$10$^{-7}$\,M$_{\sun}$\,yr$^{-1}$, and it
appears that very few AGB carbon stars {\bf ($\lesssim$5\%)} lose matter
at a rate less than $\sim$5$\times$10$^{-8}$\,M$_{\sun}$\,yr$^{-1}$.

We find that the mass loss rate and the gas expansion velocity are
relatively well correlated, but the scatter is large enough that we
may conclude that the mass loss mechanism is able to produce a wide
range of mass loss rates for a given expansion velocity.
The mass loss rate is also well correlated with the pulsational
period of the star, correlated with the stellar luminosity,
and there is a weak trend with the stellar
effective temperature, in the sense that the cooler stars tend to have
higher mass loss rates.  Also the gas expansion velocity is positively
correlated with the period and the luminosity. We conclude that the
mass loss rate increases with increased regular pulsation and/or
luminosity, and that the expansion velocity increases with mass
loss rate (for low mass loss rates) and luminosity.  The observed
trends are all supporting the common concensus that these winds are
driven by radiation pressure on dust grains, and that pulsation may
play an important role.  Somewhat surprising there appears to be no
dependence on the stellar C/O-ratio.

Our standard CSE model, assuming a single smooth expanding wind produced by
a continuous mass loss, fails to reproduce the observational data for
about 10\% of the sample stars.  Most notable among these are five
stars with detached CSEs, presumably formed during a period of highly
increased mass loss, and low present mass loss rates.  We have found
indications that the present mass loss rate is significantly higher
for the star with the youngest dCSE, and that the shell mass may
increase with shell age. Since our sample is reasonably complete, we can
estimate that the time scale between mass ejections is about 10$^5$
years, if it is a repeatable phenomenon.  An association with He-shell
flashes is favoured.  These objects have been the targets of a number
of extensive, observational and theoretical studies.

For some of our sample stars there exist enough observational
constraints to determine a combined dust parameter from the kinetic
temperature structure.  This result can be interpreted as a dust-to-gas
mass ratio which increases with mass loss rate, but changes in the dust
properties may also play a role.  We also find that this means that
the gas kinetic temperature in a carbon star CSE depends only weakly
on the mass loss rate.

The size of the CO envelope is an important parameter in the mass loss
rate determination, at least for the low mass loss rate objects. We have
used published radial CO($J$$=$$1$$\rightarrow$$0$) brightness
distributions, in combination with the radiative transfer code, to show
that the CO photodissociation calculation by Mamon et~al. \cite*{Mamon88}
gives reasonably accurate results.  In a few cases the observed radial
brightness distributions are clearly different than the model results,
suggesting deviations from a simple $r^{-2}$ density law, and hence
time-variable mass loss.

We estimate that carbon stars, of the type studied here,
return on the order of 0.05\,M$_{\sun}$\,yr$^{-1}$  of gas to the ISM making
them marginally important for the gas cycle in galaxies.
More extreme carbon stars may contribute an order of magnitude more.
However, they are probably not important as regards the origin of
carbon.

\begin{acknowledgements}
We are grateful to Dr.~F.~Kerschbaum for generously providing
          estimates of some of the input parameters to the CO modelling,
	 to Dr.  R. Liseau for useful comments,  and to an 
       anonymous referee
	 for constructive criticism which lead to an improved paper. Financial
	 support from the Swedish Natural Science Research Council (NFR) is
	 gratefully acknowledged.
\end{acknowledgements}

\bibliographystyle{aabib99}

\end{document}